\documentclass[preprint,showpacs,prb]{revtex4}
\usepackage{amssymb}
\usepackage{graphicx}
\usepackage{amsmath}
\usepackage{bm}
\usepackage{verbatim}

\begin{document}

\title{Vortex clusters and multiquanta flux lattices in thin films of anisotropic superconductors}
\author{A.~V.~Samokhvalov$^{(1)}$, D.~A.~Savinov$^{(1)}$, A.~S.~Mel'nikov$^{(1)}$,
A.~I.~Buzdin$^{(2)}$} \affiliation{$^{(1)}$ Institute for Physics
of Microstructures, Russian Academy
of Sciences, 603950 Nizhny Novgorod, GSP-105, Russia\\
$^{(2)}$ Institut Universitaire de France and Universite Bordeaux
I, France\\} \pacs{}

\begin{abstract}
The distinctive features of equilibrium vortex structures in thin
films of anisotropic superconductors in tilted magnetic fields are
studied for the limits of moderate and strong anisotropy. The
energetically favorable shape of isolated vortex lines is found in
the framework of two particular models describing these limiting
cases: London theory with an anisotropic mass tensor and
London-type model for a  stack of Josephson--decoupled
superconducting layers. The increase of the field tilting is shown
to result in qualitative changes in the vortex--vortex interaction
potential: the balance between long--range attractive and
repulsive forces occurs to be responsible for a formation of a
minimum of the interaction potential vs the intervortex distance.
This minimum appears to exist only for a certain restricted range
of the vortex tilting angles which shrinks with the decrease of
the system anisotropy parameter. Tilted vortices with such unusual
interaction potential form clusters with the size depending on the
field tilting angle and film thickness or/and can arrange into
multiquanta flux lattice. The magnetic flux through the unit cells
of the corresponding flux line lattices equals to an integer
number $M$ of flux quanta. Thus, the increase in the field tilting
should be accompanied by the series of the phase transitions
between the vortex lattices with different $M$.
\end{abstract}

\maketitle


\section{Introduction}

According to a standard picture of the mixed state in bulk type-II superconductors
the Abrikosov vortices penetrating the homogeneous sample form a periodic
arrangement called a flux lattice
\cite{Abrikosov-Fund}.
The magnetic flux through the unit cell of such flux line lattice equals
to the flux quantum $\phi_0=\pi \hbar c/e$: we have one
vortex per unit cell.
There are a few examples of rather exotic superconducting systems which may
provide a possibility to observe a different vortex lattice periodicity,
namely the structures with
more than one vortices per unit cell.
In particular, the phase transitions to such multiquanta flux lattices can occur,
e.g., for superconductors with unconventional pairing
\cite{Joynt-rmp02,Melnikov-zetf92}
or 2D Fulde-Ferrell-Larkin-Ovchinnikov superconductors
\cite{Houzet-el00}.

The goal of this work is to suggest an alternative scenario of the
phase transitions between the flux structures with different
number of vortices per unit cell which can be realized in thin
films of anisotropic superconductors. The underlying physical
mechanism for this scenario arises from the interplay between the
long range attraction and repulsion between tilted vortex lines in
thin films discussed recently in Ref.~\onlinecite{Buzdin_0}. The
unusual attractive part of the vortex--vortex interaction
potential is known to be a distinctive feature of anisotropic
superconductors and the value of the attractive force is
controlled by the tilting angle of the vortex line with respect to
the anisotropy axis \cite{Buzdin-JETPL90,Grishin,Kogan-prb90}. The
origin of the long range intervortex repulsion in thin films has
been analyzed in the pioneering work \cite{Pearl} by Pearl in
1964. This repulsion force always overcomes the attraction at
rather large distances because of the different power decay laws
of these contributions. Note that, of course, the short range
interaction between vortices is also repulsive. Finally, this
balance between the repulsion and attraction can result in the
formation of the nonmonotonic interaction potential $U(R)$ vs the
intervortex distance $R$. Increasing the vortex tilting angle we
first strengthen the attraction force between vortices  and, thus,
the minimum in the vortex interaction potential can appear only
for rather large tilting angles when the attraction overcomes the
Pearl's repulsion. This minimum shifts towards the larger
intervortex distances with the further increase in the tilting
angle and, finally, at rather large distances the attraction
appears to be suppressed due to the exponential screening effect.
As a consequence, the minimum in the interaction potential exists
only for a certain restricted range of the vortex tilting angles
which shrinks with the decrease of the system anisotropy
parameter. The appearance of a minimum in the interaction
potential points to the possibility to get a bound vortex pair (or
even the clusters with higher vorticities) for a certain range of
vortex tilting angles. For a flux line lattice such vortex--vortex
interaction potential can cause an instability with respect to the
unit cell doubling, i.e. the phase transition to the multiquanta
vortex lattices.

In this paper we use two theoretical approaches to describe the
peculiarities of the intervortex interaction and resulting
formation of clusters and multiquanta lattices. One of them is a
standard London model accounting for an anisotropic mass tensor
which is adequate for the superconductors with moderate
anisotropy. This approach assumes that the superconducting
coherence length in all directions exceeds the distance between
the atomic layers and obviously breaks down in the limit of strong
anisotropy, i.e., for Josephson--coupled layered structures. In
the latter case we choose to apply another phenomenological model,
namely the so--called Lowrence--Doniach theory
\cite{Lowrence-Doniach}. For rather small intervortex distances
this theory can be simplified neglecting the effects of weak
interlayer Josephson coupling. This approach of
Josephson--decoupled superconducting layers is known to be useful
in studies of the vortex--lattice structure at low fields
\cite{buzdin-feinberg,Clem-prb91}.

Considering thin film samples in tilted magnetic fields we do not
restrict ourselves by the case of only straight vortex lines and
study the problem of the energetically favorable vortex line shape
in the presence of the inhomogeneous supercurrent screening the
field component $\mathbf{H}_\parallel$ parallel to the film plane.
Previously this problem has been addressed in
Ref.~\onlinecite{Brandt-prb93} for rather small deviations of the
vortex line from the direction normal to the film plane. Such
approximation is obviously valid only for the $|\, \mathbf{H}_\parallel |$ values
much smaller than the critical field $H_{c1}^{(0)}$ of the penetration
of vortices parallel to the film plane. For anisotropic London
model this analysis of Ref.~\onlinecite{Brandt-prb93} has been
previously generalized for the case of a strongly distorted vortex
line (see Ref.~\onlinecite{Martynovich-JETP94}). For the sake of
completeness we present here the calculations of the shape of an
isolated vortex line for arbitrary fields
$|\, \mathbf{H}_\parallel | < H_{c1}^{(0)}$
within both theoretical models describing the limits of strong and
moderate anisotropy. As a next step, we calculate the
vortex-vortex interaction potential for such strongly deformed
vortex lines. Further analysis in the paper includes the
calculations of energy of finite size vortex clusters as well as
the energy of vortex lattices with different number of vortices
per unit cell.

Experimentally  the visualization of unconventional vortex
arrangements could be carried out by a number of methods which
provided convincing evidence for the existence of vortex chains
in bulk anisotropic superconductors caused by the intervortex
attraction phenomenon  (such as the decoration technique in
$YBa_{2}Cu_{3}O_{7}$ \cite{gavamel}, scanning-tunneling microscopy
in $NbSe_{2}$  \cite{Hess}, scanning Hall-probe \cite{Grigorenko-Nature01}
and Lorentz microscopy measurements
in $YBa_{2}Cu_{3}O_{7}$ \cite{Matsuda-Science01,Tonomura}).

The paper is organized as follows. In Sec.~II we find the
energetically favorable shape of an isolated vortex line. In
Sec.~III we calculate the vortex--vortex interaction potential and
prove the existence of a potential minimum for a certain range of
field tilting angles and parameters. The Sec.~IV is devoted to the
calculation of energy of vortex clusters. Finally, in Sec.~V we
present our analysis of the phase transition between the vortex
lattices with one and two flux quanta per unit cell. The results
are summarized in Sec.~VI. Some of the calculation details are
presented in the Appendices~\ref{Apx-A} and \ref{Apx-B}.

\section{Energetically favorable shape of an isolated vortex line}

\subsection{Vortex line in a finite stack of thin superconducting layers}

We start our study of the distinctive features of equilibrium
vortex structures in thin films of anisotropic superconductors
with the consideration of the vortex line shape in the layered
systems. Let us consider a finite stack of $N$ superconducting
(SC) layers. Vortex line of an arbitrary shape pierces the film
and can be viewed as a string of 2D pancake vortices: each of
these pancakes is centered at the point $\mathbf{r}_n = x_n
\mathbf{x}_0 + y_n \mathbf{y}_0$ in the $n$-th layer. Within the
model of the stack of Josephson--decoupled SC layers, pancakes can
interact with each other only via magnetic fields. We denote the
interlayer spacing as $s$ and consider each of the $N$ layers as a
thin film with the thickness $d$  much less than the London
penetration depth $\lambda$.  General equation for the vector
potential ${\bf A}$ distribution in such system reads
\begin{equation}\label{eq:1}
{\rm rot}\,{\rm rot}\,{\bf A} = 
\frac{4\pi}{c} \sum\limits_{n,\,m =1}^{N}
{\bf J}_n^m ({\bf r}) \, \delta(z - z_n)\,,
\end{equation}
where $\Lambda = \lambda^2 / d$ is the effective penetration depth
in a superconducting film of a vanishing thickness $d$, each
$n-$th SC layer coincides with the plane $z = z_n = n s$ ($1 \le n
\le N$), the sheet current at the $n-$th layer created by the pancake at
$m-$th layer takes the form
\begin{equation}\label{eq:2}
    \mathbf{J}_n^m(\mathbf{r})=\frac{c}{4 \pi \Lambda}
        \left[\, \mathbf{\Phi}(\mathbf{r}-\mathbf{r}_m)\,\delta_{n m}
              -\mathbf{A}^m(\mathbf{r},z_n) \,\right]\ ,
\end{equation}
$\mathbf{A}^m(\mathbf{r},z)$ is the vector potential induced by
the only pancake vortex located in the $m-$th layer (Fig.~1).
The vector $\mathbf{\Phi}(\mathbf{r})$ in the Eq.~(\ref{eq:2}) is given by the expression
\begin{equation}\label{eq:3}
   \mathbf{\Phi}(\mathbf{r})
   = \frac{\Phi_0}{2\pi} %
     \frac{\left[ \mathbf{z}_0 \times \mathbf{r} \right] }
          {\mathbf{r}^2} \,,
\end{equation}
and $\phi_0 = \pi \hbar c / e$ is the flux quantum.
For the layered system without Josephson coupling a general expression
for the free energy  can be written in the form:
\begin{equation}\label{eq:6c}
    F = \frac{1}{8\pi}\int dV
       \left[ {\left({\rm rot}\,\mathbf{A}\right)}^2
        + \left(\frac{4\pi}{c}\right)^2 \Lambda
                    \sum\limits_n \mathbf{J}_n^2(\mathbf{r})\, \delta(z-z_n)
                    \right]\,.
\end{equation}
where the total vector potential $\mathbf{A}(\mathbf{r},z)$
and the sheet current in the $n-$th layer $\mathbf{J}_n(\mathbf{r})$,
produced by an arbitrary vortex line are the sum of the contributions induced by
all 2D pancakes:
$$
\mathbf{A}(\mathbf{r},z) = \sum\limits_{m=1}^N \mathbf{A}^m(\mathbf{r},z)\,,
\qquad
\mathbf{J}_n(\mathbf{r}) = \sum\limits_{m=1}^N \mathbf{J}_n^m(\mathbf{r})\,.
$$

To find the magnetic vector potential ${\bf A}^m({\bf r},z)$ we
adopt an approach similar to that in
Refs.~\onlinecite{pudikov,Pe-Clem-prb97}.
Between the SC layers the vector potential $\mathbf{A}^m$ is
described by the Laplace equation
\begin{equation}\label{eq:4}
    \triangle\,\mathbf{A}^m(\mathbf{r}, z)= 0 \ .
\end{equation}
For the gauge $A_z^m = 0$ the vector potential has only the
in-plane components $\mathbf{A}^m = (A_x^m, A_y^m)$, where
\begin{equation}\label{eq:5}
    \mathbf{A}^m(\mathbf{r}, z)=\frac{1}{(2\pi)^3}
        \int\, d\mathbf{q}\, \mathrm{e}^{i \mathbf{q}\,\mathbf{r}}
        \mathbf{A}^m_q U^m(\mathbf{q}, z)\,,
\end{equation}
and the function $U^m(\mathbf{q}, z)$ can be written as
\begin{equation}\label{eq:6}
U^m(\mathbf{q}, z)=\left\{
        \begin{array}{c}
            \left[\, \alpha_n^m\,\sinh q(z_{n+1}-z) %
                +\alpha_{n+1}^m\,\sinh q(z-z_n)\,\right] / \sinh(q s)\,, \\
            \qquad z_n < z < z_{n+1},\; n=1\ldots N-1 \,, \\
            \\
            \alpha_N^m\,\exp\left(-q(z-z_N)\right)\,,\quad z \ge z_N \,, \\
            \\
            \alpha_1^m\,\exp\left(q(z-z_1)\right)\,,\quad z \le z_1 \,.
        \end{array} \right.
\end{equation}
Taking the Fourier transform of Eq.~(\ref{eq:2}) we find:
\begin{equation}\label{eq:7}
    \mathbf{J}_n^m(\mathbf{q})=\frac{c}{4 \pi \Lambda}
        \left[ \mathbf{\Phi}(\mathbf{q})\,
               \mathrm{e}^{i \mathbf{q} \mathbf{r}_m}\,\delta_{nm}
              -\mathbf{A}_q^m\, \alpha_n^m(\mathbf{q})\, \right],
\end{equation}
where
\begin{equation}\label{eq:8}
   \mathbf{\Phi}(\mathbf{q}) = - i \phi_0
       \frac{\left[ \mathbf{z}_0 \times \mathbf{q} \right]}{q^2} \
       .
\end{equation}
The sheet current density $\mathbf{J}_n^m$ results in the
discontinuity of the in-plane component of the magnetic field
$\mathbf{B}_\|^m$ across the $n$ layer:
\begin{equation}\label{eq:9}
    \frac{4\pi}{c}\, \mathbf{J}_n^m =
        \mathbf{z}_0 \times \left[ \mathbf{B}_\|^m(\mathbf{r},z_n+0)
-\mathbf{B}_\|^m(\mathbf{r},z_n-0) \right] =
        \mathbf{z}_0 \times \left.\left[ \mathbf{z}_0 \times \frac{\partial \mathbf{A}^m}{\partial z}
                            \right]\right|_{z_n-0}^{z_n+0} \,.
\end{equation}
Substituting the expressions (\ref{eq:5}), (\ref{eq:6}),
(\ref{eq:7}) into above condition (\ref{eq:9}) we obtain the
system of  linear equations for the coefficients $\alpha_n^m$:
\begin{eqnarray}\label{eq:10}
           & &h(q)\, \alpha_1^m - \alpha_2^m = \delta_{1 m} \,, \nonumber\\
        \nonumber\\
           &-&\alpha_{n-1}^m + g(q)\, \alpha_n^m - \alpha_{n+1}^m = \delta_{n m} \,,\quad n = 2\ldots N-1 \,, \\
        \nonumber\\
           &-&\alpha_{N-1}^m + h(q)\, \alpha_N^m  = \delta_{N m}\,. \nonumber
\end{eqnarray}
Here we introduce two new functions which depend on the wave number $q$:
$$
    g(q) = 2 \cosh(q s) + \sinh(q s) / \Lambda q\,, \qquad
    h(q) = \cosh(q s) + (1 + 1 / \Lambda q)\sinh(q s)\,.
$$
The solution of the linear system (\ref{eq:10}) and the
Eqs.~(\ref{eq:5}), (\ref{eq:6}) define the distribution of the
vector potential $\mathbf{A}^m(\mathbf{r},z)$ which is created by a single
pancake vortex positioned in the $m-$th layer.

Without the in-plane external magnetic field $\mathbf{H}_\|$ the
relative displacement of the pancakes in the different layers is
absent: $\mathbf{R}_{mk} = \mathbf{r}_m - \mathbf{r}_k = 0$ (i.e.
the pancakes form a vertical stack). A rather small magnetic field
$\mathbf{H}_\|= H_a \mathbf{y}_0$ induces  a screening Meissner
current $\mathbf{J}_n^M = J_n^M \mathbf{x}_0$ in each $n-$th
layer. Lorentz forces arising from these currents will move the
pancakes from their initial positions. Taking into account the
Eq.~(\ref{eq:9}), we find the following system of linear equations
describing the distribution of the magnetic field  screened by the
layered structure:
\begin{eqnarray}\label{eq:11}
            & &\left( 2 + \frac{s}{\Lambda} \right) H_1 - H_2 = H_a\,, \nonumber \\
            &-&H_{n-1} + \left( 2 + \frac{s}{\Lambda} \right) H_n - H_{n+1} =0 \,,
                \quad n = 2\ldots N-2 \,, \\
            &-&H_{N-2} + \left( 2 + \frac{s}{\Lambda} \right) H_{N-1} = H_a\,. \nonumber
\end{eqnarray}
Here $\mathbf{H}_n= H_n \mathbf{y}_0$ is the magnetic field value
between the $n-$th and $(n+1)-$th layers. The distribution of the
Meissner screening currents in the layers takes the form:
\begin{equation}\label{eq:12}
    J_1^M = \frac{c}{4\pi} \left( H_1 - H_a \right)\,, \quad
    J_n^M = \frac{c}{4\pi} \left( H_n - H_{n-1} \right)\,,\:n = 2\ldots N-2\,, \quad
    J_N^M = \frac{c}{4\pi} \left( H_a - H_{N-1} \right)\,.
\end{equation}
The resulting Lorentz forces $\mathbf{F}_n^M$ acting on the
pancakes can be written as follows:
\begin{equation}\label{eq:13}
    \mathbf{F}_n^M = ( \phi_0 / \, c ) \left[ \mathbf{J}_n^M \times \mathbf{z}_0 \right]
     = ( \phi_0 / \, c ) J_n^M \mathbf{y}_0\,.
\end{equation}
The interaction forces between the pancakes can be found using the
expression (\ref{eq:7}) for the sheet current $\mathbf{J}_k^m$
generated by the pancake positioned in the $m-$th layer:
\begin{equation}\label{eq:14}
        \mathbf{F}_k^m = (\phi_0 / c )
 \left[\, \mathbf{J}_k^m \times \mathbf{z}_0\, \right]
=
       \frac{\phi_0^2}{8\pi^2\Lambda\lambda_{ab}} \left\{\, %
               \frac{1}{R_{mk}}\,\delta_{mk}
              - \int\limits_0^\infty d q\, J_1(q R_{mk})\,
                \frac{ \alpha_k^m(q)\,g(q)}{Z(q)}\,
            \right\}\, \frac{\mathbf{R}_{mk}}{R_{mk}}\,,
\end{equation}
where $J_1(\zeta)$ is the first-order Bessel function of the first
kind, $\lambda_{ab}^2 = \Lambda s = \lambda^2 s/\, d $ is the
 penetration depth for the in-plane currents, and
$$
    Z(q) = 1 + 2 q \Lambda / \tanh(q s)\,.
$$

In order to find the equilibrium form of the vortex line in a
finite stack of $N$ superconducting layers  under the influence of
the in-plane external magnetic field $\mathbf{H}_\|$, we consider
the relaxation of the set of the pancakes towards the equilibrium
positions within the simplest version of the dynamic theory:
\begin{equation}\label{eq:16}
    \eta\, \frac{d \mathbf{r}_n}{d t} =
        \sum\limits_{m \ne n} \mathbf{F}_n^m + \mathbf{F}_n^M \,,
\end{equation}
where $\eta$ is the viscous drag coefficient. Considering the
vortex line consisting of $N=31$ pancakes we start from the
initial configuration of pancakes arranged in a straight vortex
line parallel  to the $z$ direction (see Fig.~2). As the system
approaches its final force-balanced (equilibrium) configuration,
the velocities of all pancake motions should vanish:
$$
    \lim\limits_{t \to \infty} \frac{d \mathbf{r}_n}{d t} = 0,
        \qquad 1 \le n \le N\,.
$$
In Fig.~2 we illustrate the evolution of the pancake
configurations for two  values of the applied in--plane magnetic
field $\mathbf{H}_\|$ and for two different numbers of layers:
$N=31$ (Fig.~2a, 2b) and $N=11$ (Fig.~2c, 2d). The forces
$\mathbf{F}_n^M$ caused by the Meissner currents rotate and bend
the vortex line. For rather small applied field values this
rotation and bending result in the formation of a certain stable
configuration (see Fig.~2a, 2c). For the fields exceeding a
certain critical value $H^*$ we do not find such stable pancake
arrangement. The vortex line splits into two segments which move
in opposite directions (see Fig.~2b, 2d). To define  the critical
value $H^*$ for the breakup of the vortex line we have carried out
the calculations of the pancake arrangements increasing the
in-plane magnetic field with the step $\delta H_a = 0.01 H_0$
($H_0 = \phi_0 / 2\pi \lambda_{ab}^2$). The stationary vortex
arrangement disappears above a certain threshold field value which
is taken as a critical field $H^*$. The pancake configurations for
the both cases $N=31$ and $N=11$ are qualitatively similar
 but the values of the
critical field $H^*$ for $N=31$ and $N=11$ are different. With a
decrease in the number of layers $N$ (film thickness) the value of
the critical field $H^*$ grows: $H^* = 0.21 H_0$ for $N=31$ and
$H^* = 0.38 H_0$ for $N=11$. In fact in layered superconductors
with very weak interlayer coupling the Josephson vortices will
appear at much lower field $H_a \sim H_{c1}^{(0)} \ll H^*$. As a
result, at tilted magnetic field, crossing lattice of pancakes,
forming Abrikosov vortices, and Josephson vortices exist rather
than a lattice of tilted vortex stacks
\cite{Grigorenko-Nature01,Bulaevskii-prb92,Koshelev-prl99}. The
interaction between pancakes and in-plane field in the form of
Josephson vortices produces zigzag deformation of the stack of the
pancakes \cite{Bulaevskii-prb96}. This deformation is responsible
for a long range attraction between such stacks
\cite{Buzdin-Baladie-prl02} which is quite similar to the case of
considered in the present work.

\subsection{Vortex line within anisotropic London model}

We proceed now with the consideration of the vortex line shape in
an anisotropic film which is characterized by the London
penetration depths $\lambda_{ab}$ and $\lambda_c$ for currents
flowing parallel and perpendicular to the $ab$ plane,
respectively. We consider the case of uniaxial anisotropy which
can be described by a dimensionless anisotropic mass tensor
$m_{ij}=m_0(\delta_{ij}+(\Gamma^2-1)\,c_ic_j)$, where
$\Gamma=\lambda_c/\lambda_{ab}$ is the anisotropy parameter and
${\bf c}$ is the anisotropy axis. We choose the $z$ axis of the
coordinate system perpendicular to the film surface. In the
parallel to the film plane direction we apply a certain magnetic
field ${\bf H}_{\parallel}={\bf y}_0H_a$ which is screened inside
the superconducting film.

We consider first a typical geometry when the $\bf{c}$ axis is
chosen along the direction normal to the film plane.
In such geometry the vortex line is parallel to the plane $(y,z)$
and can be parameterized by a single valued function $y=y(z)$.
An appropriate thermodynamic potential for determination
of the energetically favorable form of the vortex line takes the
form
\begin{eqnarray}
G=F_v-\frac{\Phi_0}{4\pi}\int_{-D/2}^{D/2}{dz\left(1-\frac{\cosh(z/\lambda_{ab})}
{\cosh(D/2\lambda_{ab})}\right)y'(z)H_a} \ , \label{anis general
gibbs}
\end{eqnarray}
where $D$ is the film thickness.  The first term, $F_v$, is the
Ginzburg-Landau free energy of the curved vortex line, and the
second term corresponds to the work of Lorentz force acting on the
flux line and distorting this line in the presence of screening
currents induced by the external magnetic-field component $H_a$
parallel to the film plane. To simplify the $F_v$ expression we
consider a strong type-II superconducting material with a large
ratio of the London penetration depths and coherence lengths. In
this case the main contribution to the vortex line energy is
determined by the energy of supercurrents
${\bf j}_v=c\,\mathrm{rot} \mathbf{B}_v/4\pi$ flowing around the vortex core
\begin{equation}
F_v\simeq\frac{\lambda_{ab}^{2}}{8\pi}\int{dV\,
\mathrm{rot}\mathbf{B}_v(\widehat{\mu}\,\mathrm{rot}\mathbf{B}_v)}\simeq
\frac{\Phi_0^{2}}{32\pi^3\lambda_{ab}^2}\int{dV(\widehat{\mu}^{-1}\nabla\theta_v,\nabla\theta_v})
\ , \label{anis free energy}
\end{equation}
where $\widehat{\mu}=\widehat{m}/m_0$, and $\theta_v$ is the order
parameter phase distribution around the vortex line. The above
expression for the free energy reveals a logarithmic divergence
which should be cut--off at both the small and large spatial
length scales. The lower cut--off length is naturally equal to the
characteristic size $r_c$ of the vortex core which is of the order
of the coherence length. Of course, in anisotropic case one should
introduce two different coherence lengths $\xi_{ab}$ and $\xi_c$
in the $ab$ plane and along the $c$ axis, respectively. The
resulting core size and lower cut--off length  $r_c$ for a certain
element of the tilted flux line will, thus, depend on both
$\xi_{ab}$ and $\xi_c$ lengths as well as on the local tilting
angle of the vortex line. The upper cut--off length strongly
depends on the ratio of the film thickness to the London
penetration depth. For rather thick films $d\gg\lambda_{ab}$ this
cut--off length $L_c$ is determined by a certain combination of
the screening lengths $\lambda_{ab}$ and $\lambda_c$ (see, e.g.,
Ref.~\onlinecite{klemm}).
For a thin film with $d\ll\lambda_{ab}$ one can separate two
energy contributions: (i) the contribution coming from the region
of the size $\sim d$ around the curved vortex line and providing
the logarithmic term in the free energy with the upper cut--off
length $L_c\simeq d$; (ii) the logarithmic contribution $\propto
\ln(\lambda_{ab}^2/d^2)$ coming from the larger distances $\rho>d$
which weakly depends on the vortex line shape. To sum up, the part
of the vortex line energy which depends on its shape can be
approximately written in the form:
\begin{equation}
 \delta F_v\simeq
\frac{\Phi_0^{2}}{16\pi^2\Gamma\lambda_{ab}^2}
\ln\left(L_c/r_c\right)\int_{-D/2}^{D/2}dz\sqrt{\Gamma^2+y'^2(z)}
\ .
\end{equation}
Note that we neglect here the weak dependence of the logarithmic
factor on the vortex line curvature and local orientation. Within
such approximation  we consider the vortex line as a thin elastic
string which is, of course, valid provided the characteristic
scale of the string bending is larger than the upper cut--off
length $L_c$.

The condition of the zero first variation of the Gibbs functional
gives us the following equation
\begin{equation}
y'(z)=\frac{\Gamma b(z)}{\sqrt{1- b^2(z)}} \label{anis forma d-arb} \ ,
\end{equation} where
\[
b(z)=
\frac{H_a}{H_{ab}}\times\bigg(1-\frac{\cosh(z/\lambda_{ab})}{\cosh(D/2\lambda_{ab})}\bigg)
\ , \
H_{ab}=\frac{\Phi_0}{4\pi\Gamma\lambda_{ab}^{2}}\ln(L_c/r_c) \ .
\]
The equation \eqref{anis forma d-arb} is valid for magnetic fields
which do not exceed  the critical field of the penetration of
vortices parallel to the film plane
\begin{eqnarray}
|H_a|<H_{c1}^{(0)}=H_{ab}\frac{\cosh(D/2\lambda_{ab})}{\cosh(D/2\lambda_{ab})-1}
\ .  \label{Bc1}
\end{eqnarray}
Thus, analogously to the case of a stack of decoupled layers the
stable curved vortex lines can exist only for rather small
magnetic fields below the critical field $H_{c1}^{(0)}$
which corresponds to the penetration of a vortex parallel to the film plane.
%
%
Note that in the limit $H_a\ll H_{ab}$ one can obtain the result
found previously in Ref.~\onlinecite{Brandt-prb93}:
\begin{equation}
\nonumber y'(z)\simeq \frac{\Gamma
H_a}{H_{ab}}\bigg(1-\frac{\cosh(z/\lambda_{ab})}{\cosh(D/2\lambda_{ab})}\bigg)
\ ,\quad y(z)\simeq\frac{\Gamma
H_a}{H_{ab}}\bigg(z-\lambda_{ab}\frac{\sinh(z/\lambda_{ab})}{\cosh(D/2\lambda_{ab})}\bigg)
\ .
\end{equation}
Typical shape of a bent vortex line calculated from Eq.~(\ref{anis forma d-arb}) is shown in Fig.~3a.

The above expressions can be easily generalized for an arbitrary
angle $\chi$ between the anisotropy axis ${\bf c}$ and the
direction normal to the film plane.
We restrict ourselves to the case when the axis ${\bf c}$ is parallel
to the plane $(y,z)$ and vortex line can be parameterized by a function $y=y(z)$
as before.
In this case the part  of the free energy \eqref{anis free energy}
depending on the vortex line shape takes
the form:
\begin{eqnarray}
\nonumber \delta F_v\simeq
\frac{\Phi_0^2}{16\pi^2\Gamma\lambda^2_{ab}}
\ln(L_c/r_c)\int_{-D/2}^{D/2} dz\sqrt{1+y'^2(z)}\sqrt{\rm sin^2 \
\theta(z)+\Gamma^2\rm cos^2 \ \theta(z)} \label{anis free energy
3} \ ,
\end{eqnarray}
where $\tan\bigg[\theta(z)+\chi\bigg]=y'(z)$. Thus we find the
following equation describing the vortex line shape:
\begin{equation}
y'(z)=\frac{t(1-\Gamma^2)}{1+t^2\Gamma^2}+\frac{\Gamma{N(z)(1+t^2)}}{(1+t^2\Gamma^2)\sqrt{1+t^2
\Gamma^2-N^2(z)}} \ , \label{aniz shape}
\end{equation}
where
\[\nonumber N(z)=\frac{H_a}{H_{ab}}\sqrt{1+t^2}\times\bigg(1-\frac{\rm
cosh(z/\lambda_{ab})}{\rm cosh(D/2\lambda_{ab})}\bigg) \ , \
t=\tan\chi \ .
\]
Note that the equation \eqref{aniz shape} is valid in the field
range
\begin{equation}
|H_a|<H_{c1}^{(\chi)}=\frac{H_{ab}\cosh(D/2\lambda_{ab})}
{\cosh(D/2\lambda_{ab})-1}\sqrt{\frac{1+t^2\Gamma^2}{1+t^2}} \
.\label{field}
\end{equation}
The critical field $H_{c1}^{(\chi)}$ corresponds to  the
penetration of a vortex parallel to the film plane. Typical plots
illustrating the numerical solution of the equation \eqref{aniz
shape} are shown in the Fig.~3b
 for different orientations of the
applied magnetic field. Note an important difference between the
opposite directions of the magnetic field $H_a$: for $H_a>0$ the
vortex line consists of segments tilted in opposite directions
with respect to the $z$ axis.

\section{Vortex--vortex interaction potential}

In this section we derive general expressions for the interaction
energy between two vortices in a thin film of an anisotropic
superconductor taking into account both long range attraction and
repulsion phenomena. We study both the limits of strong and
moderate anisotropy for a wide range of vortex tilting angles. The
shape of the interacting vortex lines is assumed to be fixed and
not affected by the vortex--vortex interaction potential.
Certainly, such assumption is valid only in the limit of rather
larger distances between the vortex lines when the effect of
interaction on the vortex shape can be viewed as a small
perturbation.

\subsection{Interaction potential of two tilted stacks of pancakes}

In this section we consider the interaction between two vortex
lines consisting of pancake vortices. For each vortex the pancake
centers are assumed to be positioned along the straight line
tilted at the angle $\gamma$ with respect to the anisotropy axis
$\mathbf{c}$ ($z$ axis). We restrict ourselves to the case of
parallel vortex lines shifted by a certain vector $\mathbf{R}$ in
the plane of the layers. Using the gauge
$\mathrm{div}{\mathbf{A}}=0$ and the Fourier transform
\begin{equation}\label{eq:1c}
\mathbf{A}(\mathbf{q},k) =\int d^2 \mathbf{r}\, dz\,
    \mathrm{e}^{i \mathbf{q} \mathbf{r} + i k z} \mathbf{A}(\mathbf{r},z)\,,
\end{equation}
\begin{equation}\label{eq:1c1}
\mathbf{A}_n(\mathbf{q}) = \int d^2 \mathbf{r}\,
    \mathrm{e}^{i \mathbf{q} \mathbf{r}} \mathbf{A}(\mathbf{r},z_n)\,,
\qquad
\mathbf{J}_n(\mathbf{q}) = \int d^2 \mathbf{r}\,
    \mathrm{e}^{i \mathbf{q} \mathbf{r}} \mathbf{J}_n(\mathbf{r})\,,
\end{equation}
one can rewrite the basic equation (\ref{eq:1})
in the momentum representation as follows:
\begin{equation}\label{eq:2c}
    \left( q^2 + k^2 \right)\,\mathbf{A}(\mathbf{q}, k) =
    \frac{1}{\Lambda} \sum\limits_n\,
        \left( \mathbf{\Phi}_n(\mathbf{q})-\mathbf{A}_n(\mathbf{q})\,
        \right)\mathrm{e}^{i k n s}\ ,
\end{equation}
where $\mathbf{\Phi}_n(\mathbf{q}) =
\mathbf{\Phi}(\mathbf{q})\,\mathrm{e}^{i \mathbf{q}
\mathbf{r}_n}$.
Taking  account of the relation
$$
    2\pi\mathbf{A}_n(\mathbf{q}) =
    \int dk\, \mathrm{e}^{i k z_n} \mathbf{A}(\mathbf{q}, k)\,,
$$
we obtain from (\ref{eq:2c}) the following equations for the
Fourier components of the vector potential
$\mathbf{A}_n(\mathbf{q})$:
\begin{equation}\label{eq:3c}
    2 q \Lambda\, \mathbf{A}_n =
        \sum\limits_m
        \left( \mathrm{e}^{i \mathbf{q} \mathbf{r}_m}
               \mathbf{\Phi}(\mathbf{q}) - \mathbf{A}_m
        \right) \mathrm{e}^{-| n - m | q s}\,.
\end{equation}
These equations can be reduced to the scalar form
\begin{equation}\label{eq:4c}
    f_n + \frac{1}{2 q \Lambda} \sum\limits_m
        \mathbf{e}^{-| n - m | q s}\, f_m
    =  \mathbf{e}^{i \mathbf{q} \mathbf{r}_n} \ ,
\end{equation}
where we introduce the new functions $f_n(\mathbf{q})$:
\begin{equation}\label{eq:5c}
    \mathbf{J}_n (\mathbf{q}) = \frac{c}{4\pi\Lambda}
        \left(\,
               \mathbf{\Phi}_n(\mathbf{q}) - \mathbf{A}_n(\mathbf{q})\,
        \right)
        = \frac{c}{4\pi\Lambda}\,
        \mathbf{\Phi}(\mathbf{q})\,f_n(\mathbf{q})\,.
\end{equation}
The solution of the linear system (\ref{eq:4c}) for a fixed
distribution of pancakes $\mathbf{r}_n$ determines the
distribution of the vector potential $\mathbf{A}(\mathbf{r},z)$
which is created by an arbitrary vortex line in a finite stack of
superconducting layers.
In the momentum
representation the general expression (\ref{eq:6c}) for the free
energy of the layered system without Josephson
coupling reads
\begin{equation}\label{eq:7c}
    F=\frac{1}{32\pi^3\Lambda}\, \sum_n\int d^2\mathbf{q}
        \left(\, \mathbf{\Phi}_n(\mathbf{q}) - \mathbf{A}_n(\mathbf{q})\,
        \right)\,\mathbf{ \Phi}_n(-\mathbf{q})\,.
\end{equation}
For two vortex lines we can write the total vector potential and
the total sheet current as  superpositions of contributions coming
from the first ($\mathbf{A}_n^{(1)}$, $\mathbf{J}_n^{(1)}$) and
second ($\mathbf{A}_n^{(2)}$, $\mathbf{J}_n^{(2)}$) vortices:
$$
    \mathbf{A}_n(\mathbf{q}) = \mathbf{A}_n^{(1)}(\mathbf{q})
                    +  \mathbf{A}_n^{(2)}(\mathbf{q})\,,
    \qquad
    \mathbf{J}_n(\mathbf{q}) = \mathbf{J}_n^{(1)}(\mathbf{q})
                    +  \mathbf{J}_n^{(2)}(\mathbf{q}).
$$
Calculating the interaction energy $\varepsilon_{int}$ of vortex lines
we should keep in (\ref{eq:7c}) only the terms which contain
the products of fields corresponding to different vortex lines:
\begin{equation}\label{eq:8c}
\varepsilon_{int}=\frac{1}{32\pi^3\Lambda}\, \sum_n\int d^2\mathbf{q}
    \left[\,
       \left(\,\mathbf{\Phi}_n^{(1)}(\mathbf{q}) - \mathbf{A}_n^{(1)}(\mathbf{q})\,\right)
       \mathbf{\Phi}_n^{(2)}(-\mathbf{q}) +
       \left(\,\mathbf{\Phi}_n^{(2)}(\mathbf{q}) - \mathbf{A}_n^{(2)}(\mathbf{q})\,\right)
       \mathbf{\Phi}_n^{(1)}(-\mathbf{q})\,
    \right]\,.
\end{equation}
Finally, for the particular case of two parallel vortex lines
which are shifted at the vector $\mathbf{R} = \mathbf{r}_n^{(2)} -
\mathbf{r}_n^{(1)}\: (n=1,N)$ in the $(xy)$ plane we get following
expression for the interaction energy via the scalar functions
$f_n(\mathbf{q})$:
\begin{equation}\label{eq:9c}
\varepsilon_{int}(\mathbf{R})=\frac{\phi_0^2}{16\pi^3\Lambda}\, \int \frac{d^2\mathbf{q}}{q^2}\,
        \cos(\mathbf{q} \mathbf{R})\,
        \sum_n f_n(\mathbf{q})\, \mathrm{e}^{-i \mathbf{q} \mathbf{r}_n}\, .
\end{equation}
The expression (\ref{eq:9c}) and equations (\ref{eq:4c}) determine
the interaction energy of two identically bent vortex lines.

Our further consideration in this subsection is based on two
assumptions: (i) for each vortex we choose the centers of pancakes
to be positioned along the straight line tilted at a certain angle
$\gamma$ relative to $z$ axis, and put $\mathbf{r}_{n}=n s
\tan\gamma\, \mathbf{y}_{0}$; (ii) we restrict ourselves by the
continuous limit assuming $q s \ll 1$ and $q_y s \tan\gamma \ll 1$.
In this case  the  Eqs.~(\ref{eq:4c}), (\ref{eq:9c}) can be simplified
(see Appendix~\ref{Apx-A} for details):
\begin{eqnarray}
    & &\varepsilon_{int}(\mathbf{R}) = \frac{\phi_0^2}{16\pi^3\lambda_{ab}} %
        \int\,d^2\mathbf{q}\,\cos(\mathbf{q} \mathbf{R})\, S(\mathbf{q})\,,         \label{eq:10c} \\
    & &S(\mathbf{q}) = \frac{1}{\lambda_{ab}\, q^2}\left\{ \,
        D \frac{p^2 + k^2}{1 + p^2} \right.  \nonumber \\
        & & \quad \left. + \frac{ 2(1 - k^2)
        \left[\, k (1 - p^2)\sinh L + (k^2 - p^2)( \cosh L - \cos(pL) )
                + 2 k p \sin(pL) \right] }
                        { \sqrt{q^2 + \lambda_{ab}^{-2}}\,(1 + p^2)^2
                        \left[ 2 k \cosh L + (1 + k^2)\sinh L)\right] } \right\}\,, \label{eq:11c}
\end{eqnarray}
where
\begin{equation} \label{eq:10c1} \\
L = D \sqrt{q^2 + \lambda_{ab}^{-2}}\,, \quad k = q /\sqrt{q^2 +
\lambda_{ab}^{-2}}\,, \quad p = q_{y} \tan\gamma / \sqrt{q^2 +
\lambda_{ab}^{-2}} \,,
\end{equation}
and $D = (N-1) s$ is the thickness of the superconducting film.
The first term in (\ref{eq:11c}) describes the interaction
in the bulk system, while the second term is responsible for the effect
of film boundaries.

The minimum energy configuration corresponds to the case
$R_{x}=0$. In Fig.~4 we present some typical plots of the
interaction energy $\varepsilon _{int}(R_x=0,R_y)$ vs the
distance $R_{y}=R$ for $d=3 \lambda_{ab}$ which corresponds to the
Lorentz microscopy experiments in YBCO \cite{Buzdin_0}
and Bi-2212 \cite{Tonomura} samples.
Analyzing the dependence $\varepsilon _{int}(R),$ one can
separate three contributions to the energy of vortex--vortex
interaction: (i)\ a short--range repulsion which decays
exponentially with increasing intervortex distance $R$ (for
$R>\lambda _{ab}$); (ii) an intervortex attraction which is known
to be specific for tilted vortices in anisotropic systems; this
attraction energy term decays as $R^{-2}$ and strongly depends on
the angle $\gamma $ between the vortex axis and the $\mathbf{c}$
direction; (iii) long--range (Pearl) repulsion which decays as
$R^{-1}$ and results from the surface contribution to the energy.
Note that the third term does exist even for a large sample
thickness $D$ (see Ref. \cite{Pearl2}) although in the limit $D\gg
\lambda _{ab}$ it is certainly masked by the dominant bulk contribution. At $%
R\gg \lambda _{ab}$ the short--range interaction term vanishes and
the interaction energy vs $R\ $ takes the form
\begin{equation}
\varepsilon _{int} \simeq \frac{\phi _{0}^{2}}{8\pi ^{2}}\left( -\frac{%
D_{eff}\tan ^{2}\gamma }{R^{2}}+\frac{2}{R}\right) \ ,
\label{energy-interact}
\end{equation}%
where $D_{eff}=D-2\lambda _{ab}\tanh (D/2\lambda _{ab})$ is the
effective film thickness. One can observe here an interplay
between the long-range attractive (first
term in Eq.(\ref{energy-interact})) and the repulsive (second term in Eq.(\ref%
{energy-interact})) forces. Note that the $\lambda _{ab}$ value
increases with an increase in temperature, thus, the effective
thickness decreases and the long range attraction force appears to
be suppressed with increasing temperature. For large $R$ the
energy is always positive and corresponds to the vortex repulsion
similar to the one between the pancakes in a single layer system.
With a decrease in the distance $R$ the attraction force comes
into play resulting in the change of the sign of the energy. Such
behavior points to the appearance of a minimum in the interaction
potential.
%

%
%
%
\subsection{Vortex--vortex interaction within anisotropic London model}

We now proceed with the derivation of the  expression for the
intervortex interaction energy in an anisotropic superconducting
film. We choose the anisotropy axis $\mathbf{c}$ ($z-$axis) to be
oriented perpendicular to the film plane and consider two curved
vortex lines with identical shapes found in Sec.~II B. Our further
calculations are based on general expressions derived in
Ref.~\onlinecite{Brandt-prb00} for the energy of an arbitrary
vorticity distribution in an anisotropic superconducting film of
finite thickness (see Appendix~\ref{Apx-B} for details). The
resulting interaction energy of two curved vortices shifted from
each other in the $y$ direction at a certain distance $R$ can be
presented in the form:
\begin{equation}\label{int energy}
\varepsilon_{int}=\varepsilon_0
\left(\epsilon_{int}^{vi}+\epsilon_{int}^{stray}+\epsilon_{int}^{vac}\right)
\ ,
\end{equation}
where $\varepsilon_0=\phi_0^2/16\pi^3\lambda_{ab}$, while
$\epsilon_{int}^{vi}$ , $\epsilon_{int}^{stray}$ , and
$\epsilon_{int}^{vac}$  are given by the expressions
(\ref{eq:b7}), (\ref{eq:b8}), (\ref{eq:b9}).

Considering the particular case of straight vortex lines parallel
to the plane $(yz)$ and tilted at a certain angle $\gamma $ with
respect to the $\mathbf{c}$ direction we obtain the following
expression for the interaction energy of two vortices:
\begin{eqnarray}
    & &\varepsilon_{int}(\mathbf{R}) = \frac{\phi_0^2}{16\pi^3\lambda_{ab}} %
        \int\,d^2\mathbf{q}\,\cos(\mathbf{q} \mathbf{R})\,S_\Gamma(\mathbf{q})\,, \label{straight-tilted} \\
    & &S_\Gamma(\mathbf{q}) = \frac{1}{\lambda_{ab}\, q^2} \left\{ \,
        D \left( \frac{1 + k_\Gamma^2}{1 + p_\Gamma^2} - \frac{1}{(1+q^2)\,(1+p^2)} \right)  \right. \nonumber \\
    &&\qquad \left.  + \frac{ 2(1 - k^2)\left[\, k (1 - p^2)\sinh L + (k^2 - p^2)( \cosh L - \cos(pL) )
              + 2 k p \sin(pL) \right] }
                { \sqrt{q^2 + \lambda_{ab}^{-2}}\,(1 + p^2)^2
                        \left[ 2 k \cosh L + (1 + k^2)\sinh L)\right] } \right\}\,, \label{straight-tilted2}
\end{eqnarray}
where
$$
    k_\Gamma = q \tan\gamma /\sqrt{\Gamma^2 q^2 + \lambda_{ab}^{-2}}\,, \quad
    p_\Gamma = q_{y} \tan\gamma / \sqrt{\Gamma^2 q^2 + \lambda_{ab}^{-2}}\,,
$$
and the parameters $L$, $k$ and $p$ are described by the
expressions (\ref{eq:10c1}). In the limit of strong anisotropy
($\Gamma >> 1$) the spectral function $S_\Gamma(\mathbf{q})$
(\ref{straight-tilted2}) naturally coincides with the
corresponding expressions (\ref{eq:11c}) obtained for the layered
system without Josephson coupling.

Some typical plots of the interaction energy vs the intervortex
distance for tilted vortex lines calculated using the
Eqs.~(\ref{straight-tilted}),(\ref{straight-tilted2})
are shown in Figs.~\ref{Fig:5},\ref{Fig:51}.
Analyzing the dependence $\varepsilon_{int}(R)$
one can separate three contributions to the energy of intervortex
interaction: (i)\ a short--range repulsion (for
$R\ll\lambda_{ab}\sqrt{1+\tan^2\gamma}$) which decays
exponentially with increasing intervortex distance $R$; (ii) an
intervortex attraction which comes into play for the region
$\lambda_{ab}\sqrt{1+\tan^2\gamma}<R<\Gamma\lambda_{ab}$ and
decays exponentially with the vortex--vortex distance $R$ for
$R>\Gamma\lambda_{ab}$; (iii) long--range (Pearl) repulsion which
decays as $R^{-1}$ at large distances and results from the surface
contribution to the energy. Taking the limit
$R\ll\lambda_{ab}\sqrt{1+\tan^2\gamma}$ we get
\begin{eqnarray}\nonumber \varepsilon_{int}/\varepsilon_0
    \simeq \frac{D\pi\sqrt{\Gamma^2+\tan^2\gamma}}{\Gamma}
    \ln\left(\frac{L_c}{R}\right) \ .
\end{eqnarray}
In the region $\lambda_{ab}\sqrt{1+\tan^2\gamma}<R<\Gamma\lambda _{ab}$
the short--range interaction term vanishes and the interaction
energy vs $R$ is given by the sum (\ref{energy-interact}) of
attractive and Pearl's contributions.
Similarly to the case of decoupled layers discussed above the
attractive term can result in the appearance of a minimum in the
dependence of the vortex--vortex interaction potential vs $R$.
The position of this minimum can be roughly
estimated as the boundary of the region of the short--range
repulsion: $R_{min}\simeq\lambda_{ab}\sqrt{1+\tan^2\gamma}$.
Obviously, the minimum should disappear provided
$R_{min}>\Gamma\lambda_{ab}$, i.e., when the region of the
attraction between vortices vanishes. This condition gives us the
the upper boundary on the tilting angle $\gamma$ restricting the
interval of the energy minimum existence:
\begin{equation}
\nonumber \tan ^{2}\gamma <\Gamma^2-1 \ .
\end{equation}
The lower boundary of this angular interval can be found comparing
the attractive and repulsive terms in the expression
(\ref{energy-interact}) at the distance $R_{min}$:
\begin{equation}
\nonumber \tan ^{2}\gamma >\frac{2\lambda_{ab}^2}{D_{eff}^2}
\left(1+\sqrt{1+\frac{D^2_{eff}}{\lambda_{ab}^2}}\right)\ .
\end{equation}
%
%
These analytical estimates of the angular interval are in a rough
qualitative agreement with the numerical calculations (see
Figs.~\ref{Fig:5},\ref{Fig:51}) for two values of the film
thickness $D= 3\lambda_{ab},\,10\lambda_{ab}$. Indeed,
one can see that increasing the tilting angle
we first deepen the minimum in the interaction potential and then
make it more shallow.
The 
figures confirms the deepening of the minimum with the increase in
the anisotropy parameter $\Gamma$. Our numerical calculations
demonstrate that for the film thickness $D=3\lambda_{ab}$
(Fig.~\ref{Fig:5}) the minimum of the interaction energy of two
straight tilted vortices can appear only for $\Gamma \gtrsim 14$.
Starting from $\Gamma \thickapprox 27$ the bound vortex pair
becomes energetically favorable. An increase in the film thickness
reduces the relative contribution of Pearl repulsion to the energy
of intervortex interaction $\varepsilon_{int}$. As a result
attraction of vortices takes place for smaller values of the
tilting angle $\gamma$ and anisotropy parameter $\Gamma$. Thus, in
a film with the thickness $D=10\lambda_{ab}$ (Fig.~\ref{Fig:51})
the minimum in the $\varepsilon_{int}(R)$ dependence appears for
$\Gamma \gtrsim 7$, whereas creation of the bound vortex pair
becomes energetically favorable for $\Gamma \gtrsim 9$.
%
%

As a next step, we check if the above results obtained for
straight tilted vortices remain valid for the curved vortex lines.
Our analysis of the effect of the vortex line curvature is carried
out for model vortex profiles found in Sec.~II B.
The resulting typical dependencies of the intervortex interaction potential
vs $R$ for different magnetic field values and anisotropy parameters
are shown in Figs.~\ref{Fig:6},\ref{Fig:7}.
One can clearly see that the minimum in the interaction potential
vs $R$ survives when we take account of the vortex line curvature.
Moreover the curving of the vortex line even deepens this minimum as
it is confirmed by the comparison of
energies of straight tilted and curved vortices presented in
Fig.~\ref{Fig:7}. For such comparison we choose the straight
vortex lines connecting the ends of curved vortices. We find that
for curved vortices the energy minimum exists even for smaller
anisotropy parameters than for straight vortices (i.e., the
threshold anisotropy value for $D=3\lambda_{ab}$ becomes less than
$\Gamma \approx 14$). Of course, increasing the film thickness one can
weaken the restrictions on the existence of the minimum in the
interaction potential: e.g., for $D = 10\lambda_{ab}$ the minimum
appears at $\Gamma \gtrsim 9$.

The above theoretical analysis demonstrates that
vortex--vortex attraction and the formation
of chains are possible only for the rather large tilting angles
and at low vortex concentrations, i.e., when the magnetic-field component $H_z$
perpendicular to the film plane is very weak.
In fields $H_z$ slightly above $H_{c1}$ Abrikosov vortices will form chains due to the
long range attractive interaction.
Peculiarities of penetration of such chains of tilted Abrikosov
vortices into bulk layered (anisotropic)
superconductor are well known: in the first approximation, the vortex period
in chains does not depend on applied magnetic field, while the distance between
chains changes as $1/H_z$.
The presence of vortex chains significantly modifies the magnetization curves
with respect to analogous curves for isotropic superconductors.
\cite{Buzdin-JETP90}
In the next sections we discuss additional peculiarities of intervortex interaction
specific for thin--film samples of layered (anisotropic) superconductors.

\section{Vortex clusters}

The unusual vortex-vortex interaction potential behavior discussed
in the previous section can result in unconventional vortex
structures. We start our analysis of energetically favorable
vortex structures from the problem of stability of a vortex chain.
The formation of infinite vortex chains is known to be a signature
of the intervortex attraction in  bulk anisotropic
superconductors. The long range repulsion of vortices in thin
films can destroy the infinite vortex chains. Indeed, despite of
the fact that two vortices attract each other at a certain
distance, further increase in the number of vortices arranged in a
chain can be energetically unfavorable because of the slower decay
of the repulsive force compared to the attractive one. In this
case, for rather thin samples, there appears an intriguing
 possibility to observe vortex chains of finite length, i.e., vortex molecules or clusters.
In this section we present the calculations of energies of such
vortex clusters.

As we have demonstrated above, the minimum in the
interaction potential exists for both the limits of strong and
moderate anisotropy. The vortex molecule cohesion
energy is given by the expression:
\begin{equation}\label{eq:1d}
\varepsilon_{int}^{(N)}=\sum\limits_{i>j}\varepsilon_{int}(R_{ij})\,
\end{equation}%
where $N$ is the number of vortices in the molecule, and $R_{ij}$
are the distances between $i-$th and $j-$th vortices in the chain molecule.
Shown in Figs.~\ref{Fig:8},\ref{Fig:9} are typical plots of the
interaction energy per vortex vs the intervortex distance $R$ for
equidistant vortex chains with different $N$ numbers calculated
within the model of decoupled superconducting layers and
anisotropic London theory.
The energetically favorable number of vortices in a molecule grows
as we increase the film thickness and/or the tilting angle because
of the increasing attraction term in the pair potential $\varepsilon_{int}$.
Shown in the insets of Figs.~\ref{Fig:8} are schematic pictures of
 vortex matter consisting of dimeric and trimeric molecules.
Finally, for rather thick samples with $D\gg \lambda _{ab}$ we get
a standard infinite chain structure typical for bulk systems. Note
that the formation of an infinite vortex chain may be considered in
some sense as a polymerization of the vortex molecules.
Certainly, the crossover from the vortex molecule state to the
infinite chain structure is strongly influenced by the increase in
the vortex concentration governed by the component $B_z$ of the
external magnetic field perpendicular to the film. Indeed, one can
expect such a cross-over to occur when the mean intervortex spacing
approaches the molecule size. Thus, the vortex molecule state can
appear only in a rather weak perpendicular field when its observation
can be complicated, of course, by the pinning effects.
%
%

\section{Phase transitions in vortex lattices}

Considering the effect of a finite magnetic field (i.e., a finite
concentration of vortex clusters) we restrict ourselves by the
simplest case of regular vortex arrays. For a regular vortex array
the formation of clusters corresponds to the transition with a
change in the number of vortices in the elementary lattice cell.
The mechanism underlying such transition is naturally connected
with the appearance of the minimum in the interaction potential
for a vortex pair. In this section we present our calculations of
energy of vortex lattices with different number of flux quanta per
unit cell. The possibility to get the energetically favorable
states with a few vortices per unit cell will be illustrated for a
particular intervortex interaction potential derived above for a
model of decoupled superconducting layers. The generalization of
such consideration for  anisotropic London theory is
straightforward. Note that the vortex lattice structure for bulk
anisotropic superconductors in tilted field in the framework of
London approach has been calculated in
Ref.~\onlinecite{Buzdin-PhysC91}.

Let's consider a vortex lattice characterized by the translation
vectors $\mathbf{T}= i\, \mathbf{a}_1 + j\, \mathbf{a}_2$, where
$\mathbf{a}_{1,2}$  are primitive vectors of the lattice. The
primitive cell occupies the area
$A_0=[\mathbf{a}_1\times\mathbf{a}_2]\,\cdot \mathbf{z}_0$ and is
assumed to contain $M$ vortices: $B_z A_0 = M \Phi_0$. Positions
of vortices in a cell are determined by the vectors
$\mathbf{r}_{m}$ ($m=1,M$) (see Fig.~\ref{Fig:10}).
%
%
The interaction energy per unit lattice cell can be expressed via the  vortex--vortex interaction potentials
(\ref{eq:10c}),(\ref{eq:11c}):
\begin{equation}\label{eq:1e}
\varepsilon_c (\mathbf{r}_{mk},\mathbf{T}) = \sum_{m,\,k\ne m}^M \varepsilon_{int}(\mathbf{r}_{mk})
    +\sum_{\mathbf{T} \ne 0} \sum_{m,\,k}^M
         \varepsilon_{int}(\mathbf{T}+\mathbf{r}_{mk})\,.
\end{equation}
The interaction energy (\ref{eq:1e}) depends on both the relative
positions $\mathbf{r}_{mk} = \mathbf{r}_m - \mathbf{r}_k$ of
vortices in the primitive cell and the structure of the vortex
lattice defined by the translation vectors $\mathbf{T}$. The first
term in (\ref{eq:1e}) describes the interaction energy between
vortices in the primitive cell (without the lattice contribution),
whereas the second sum takes account of the lattice effects. With
the help of the Poisson formula, one can rewrite the intervortex
interaction energy (\ref{eq:1e}) in terms of the Fourier
components
\begin{equation}\label{eq:2e}
    \varepsilon_c = \frac{\phi_0^2}{16\pi^3\lambda_{ab}}
        \left[\, \frac{4\pi^2}{A_0} \sum_\mathbf{Q} \sum_{m,\,k}^M
            S(\mathbf{Q})\,\cos(\mathbf{Q} \mathbf{r}_{mk})
            - M \int\, d^2 \mathbf{q}\, S(\mathbf{q})\, \right]\,,
\end{equation}
where the function $S(\mathbf{q})$ is determined by the
Eq.~(\ref{eq:11c}), and $\mathbf{Q}$ are the reciprocal--lattice
vectors. The sum and the integral in Eq.~(\ref{eq:2e}) diverge
both at $\mathbf{Q}=0$ and at large $\mathbf{Q}$ values. The small
$\mathbf{Q}$ divergence corresponds to the linear (in the system
size) increase in the vortex energy because of the slow $1/R$
decay of the vortex-vortex interaction potential. The large
$\mathbf{Q}$ divergence is logarithmic and is associated with the
vortex self energy.

For simplicity, we restrict ourselves to the case of an
instability with respect to the unit cell doubling and tripling,
i.e. formation of the vortex lattices with two and three flux
quanta per unit cell ($M = 2$ and $M=3$). Hereafter we consider
only the shifts of vortex sublattices along the $y$ direction and
choose the appropriate reciprocal--lattice vectors
$$
    \mathbf{Q}_{ij}= \frac{2\pi}{b}\,( i - j/4 )\,\mathbf{x}_0
          + \frac{\pi}{a}\, j\, \mathbf{y}_0\,,\qquad
        i\,, j = 0,\pm 1, \pm 2,\ldots \ ,
$$
$$
    \mathbf{Q}_{ij}= \frac{2\pi}{b}\,( i - j/6 )\,\mathbf{x}_0
          + \frac{2\pi}{3a}\, j\, \mathbf{y}_0\,,\qquad
        i\,, j = 0,\pm 1, \pm 2,\ldots
$$
for $M = 2$ and $M=3$, respectively. Here we consider only
equidistant vortex chains within the primitive cells.  Fixing the
value of the field $B_z$ we fix the unit cell area area $A_0=2a b$
for $M=2$ and $A_0=3a b$ for $M=3$. Thus, the interaction energy
(\ref{eq:2e}) depends only on two parameters: (i) $\sigma=b/a$
ratio characterizing the lattice deformation; (ii) relative
displacement $\Delta a$ of vortex sublattices along the $y$-axis
(see Fig.~\ref{Fig:10}). To exclude the divergence at $Q=0$ it is
convenient to deal with the energy difference:
\begin{equation}\label{eq:3e}
    \Delta\varepsilon_c =
       \underset{\sigma}{\rm min}\{\varepsilon_c(\sigma, \Delta a)\}
     - \underset{\sigma}{\rm min}\{\varepsilon_c(\sigma,0)\}\,.
\end{equation}
%
%
The results of our numerical calculations of this energy
difference are shown in Fig.~\ref{Fig:11}a. One can clearly observe that
changing the vortex tilting angle we obtain the minimum in the
function $\Delta\varepsilon_c (\Delta a)$ which gives us the
evidence for the phase transition in the lattice structure with
the unit cell doubling or tripling depending on the vortex tilting
angle. The multiplication of the unit cell is accompanied by the
strong change in the lattice deformation ratio $\sigma$ (see
Fig.~\ref{Fig:11}b).

\section{Conclusions}

To sum up, we suggest a scenario of the phase transitions between
the flux structures with different number of vortices per unit
cell which can be realized in thin films of anisotropic
superconductors placed in tilted magnetic fields. We demonstrate
that the vortex interaction in the films of anisotropic
superconductors placed in tilted magnetic fields is very special.
The underlying physics arises from the interplay between the long
range attraction and repulsion between tilted vortex lines. In
consequence, new and very reach types of vortex structures may
appear. They are formed from the vortex dimers, trimers, etc., and
the transition between different types of vortex structures may be
controlled by tilting of external magnetic field and/or by varying
of the temperature. Our theoretical findings are based on two
theoretical approaches:  anisotropic London model and the
London--type model of decoupled superconducting layers. Taking
account of the vortex tilt and bending we analyzed the distinctive
features of the vortex--vortex interaction potential in a wide
range of parameters and fields and demonstrated the possibility to
obtain a minimum in the vortex interaction potential vs the
intervortex distance. Further analysis in the paper included the
calculations of energy of finite size vortex clusters as well as
the energy of regular vortex arrays with different number of
vortices per unit cell. The phase transitions accompanied by the
multiplication of the primitive lattice cell appear to be possible
for dilute vortex arrays, i.e. for rather small magnetic field
component $B_z$. We believe that our theoretical predictions
concerning the unusual vortex configurations are experimentally
observable using the modern vortex imaging methods such as Lorentz
microscopy, scanning tunneling microscopy, scanning Hall-probe or
decoration technique.

\acknowledgments
We are grateful to Professor  A. Tonomura  for stimulating discussions.
This work was supported, in part, by the Russian Foundation for
Basic Research, Russian Academy of Sciences under the Program
``Quantum physics of condensed matter", Russian Agency of
Education under the Federal Program ``Scientific and educational
personnel of innovative Russia in 2009--2013", and ``Dynasty" foundation.

\appendix

\section{Interaction energy of tilted vortices: continuous limit}\label{Apx-A}

Let us evaluate the interaction energy (\ref{eq:9c}) of two tilted parallel vortex lines taking
$$
    \mathbf{r}_{n}^{(1)} = n s \tan\gamma\, \mathbf{x}_{0}\,,
    \quad
    \mathbf{r}_{n}^{(2)} = \mathbf{r}_{n}^{(1)} + \mathbf{R}
$$
and assuming  $q s \ll 1$ and $q_x s \tan\gamma \ll 1$. We
introduce a continuous coordinate $t = n s$ and continuous
function $f_\mathbf{q}(t)$. Thus, the linear system of equations
(\ref{eq:4c}) reduces to the following integral equation
\begin{equation}\label{eq:a1}
 f_\mathbf{q}(t)+ \frac{1}{2 q \lambda_{ab}^2} %
    \int\limits_{-D/2}^{D/2} dt^\prime \mathrm{e}^{-q |t - t^\prime |}
    f_\mathbf{q}(t^\prime)  =  \mathrm{e}^{i q_x t \tan\gamma }\, .
\end{equation}
The equation (\ref{eq:a1}) can be rewritten as a differential one
\begin{equation}\label{eq:a2}
     \frac{d^2 f_\mathbf{q}}{dt^2} -
        \left( \lambda_{ab}^{-2} + q^2 \right) f_\mathbf{q}(t)
        = - \left( q_x^2 \tan^2\gamma + q^2 \right) \mathrm{e}^{i q_x t \tan\gamma}\,
\end{equation}
at the interval $-D/2 < t < D/2$ with the boundary conditions
\begin{equation}\label{eq:a3}
    \left.\left(\frac{d f_\mathbf{q}}{d t} \pm q f_\mathbf{q} \right) \right|_{\pm D/2}
    = (i q_x \tan\gamma \pm q) e^{\pm i q_x D \tan\gamma /2} \ .
\end{equation}
Introducing the notations
$$
    \tau = t \sqrt{q^2 + \lambda_\|^{-2}}\,,
    \quad
    L = D \sqrt{q^2 + \lambda_\|^{-2}}\,,
    \quad
    k = q / \sqrt{q^2 + \lambda_\|^{-2}}\,,
    \quad
    p =q_x \tan\gamma /\sqrt{q^2 + \lambda_\|^{-2}}\,,
$$
one can rewrite the equation (\ref{eq:a2}) and boundary conditions
(\ref{eq:a3}) in dimensionless form
\begin{eqnarray}
    & &\frac{d^2 f_\mathbf{q}}{d \tau^2}- f_\mathbf{q}
    = -\left( p^2 + k^2 \right) \mathrm{e}^{i p \tau}\,, \label{eq:a4} \\
    & &\left.\left(\frac{d f_\mathbf{q}}{d \tau}
    \pm  k f_\mathbf{q}\right)\right|_{\pm L/2}
    =(i p \pm k) \mathrm{e}^{ip L /2}\,. \label{eq:a5}
\end{eqnarray}
The  solution of the Eq.~(\ref{eq:a4}) has the form
\begin{equation}\label{eq:a6}
    f_\mathbf{q}(\tau) = \frac{p^2+k^2}{1+p^2} \mathrm{e}^{i p \tau}
        + \left( 1 - \frac{p^2+k^2}{1+p^2} \right)
            \left( a\,\mathrm{e}^\tau + b\, \mathrm{e}^{-\tau} \right)\,,
\end{equation}
where the constants $a$ and $b$ are defined by the boundary
conditions (\ref{eq:a5}):
$$
    a = \frac{\mathrm{e}^{(i p + 1) L/2}(k + i p)(1 + k)
            + \mathrm{e}^{-(i p + 1) L/2}(k - i p)(1 - k)}
             { 2 ( 2 k \cosh L + (1 + k^2) \sinh L)}
$$
$$
    b = \frac{\mathrm{e}^{(i p - 1) L/2}(k + i p)(1 - k)
            + \mathrm{e}^{(-i p + 1) L/2}(k - i p)(1 + k)}
            {2( 2 k \cosh L + (1 + k^2) \sinh L)}\,.
$$
In the continuous limit the expression for the interaction energy
(\ref{eq:9c}) takes the form:
\begin{equation}\label{eq:a7}
    \varepsilon_{int}=\frac{\phi_0^2}{16\pi^3 \lambda_{ab}^2}
    \int\frac{d^2 \mathbf{q}}{q^2} \cos(\mathbf{q} \mathbf{R})\,
    S(\mathbf{q})\ .
\end{equation}
Here the function
$$
    S(\mathbf{q}) = \int \limits_{-D/2}^{D/2}
        d t f_\mathbf{q}(t)\,\mathrm{e}^{-i q_x t \tan\gamma}\,
$$
can be calculated analytically:
\begin{equation}\label{eq:a8}
 S(\mathbf{q}) =
    D \frac{p^2 + k^2}{1 + p^2} + \frac{2 (1 - k^2)\left[\,k(1 - p^2)\sinh L +
        (k^2 - p^2)(\cosh L - \cos(pL) ) + 2 k p \sin(pL) \right]}
        {\sqrt{q^2 + \lambda_{ab}^{-2}}\,(1 + p^2)^2\, \left[\,2 k \cosh L + (1 + k^2)\sinh L\,\right]}
        \ .
\end{equation}

\section{Interaction energy of curved vortices: anisotropic London model}
\label{Apx-B}

To calculate the vortex--vortex interaction within  anisotropic
London model we use general expressions derived in
Ref.~\onlinecite{Brandt-prb00} for the total energy $E$ of an
arbitrary arrangement of curved vortices in a superconducting film
of thickness $D$  with the $c$-axis perpendicular to the film
plane:
\begin{equation}\label{eq:b0}
E=E_{vi}+E_{stray}+E_{vac} \ ,
\end{equation}
where

\begin{eqnarray}
\nonumber
E_{vi}=\frac{\Phi_0^2}{16\pi}\int\frac{d^2k_{\bot}}{4\pi^2}\frac{1}{2D}\sum_m
\sum_{\alpha}G_{\alpha}({\bf k_{\bot}},k_m)|\nu_{\alpha}^{vi}({\bf
k_{\bot}},k_m)|^2 \ , \label{E_vi}
\end{eqnarray}

\begin{eqnarray}
\nonumber
E_{stray}=\frac{1}{8\pi}\int\frac{d^2k_{\bot}}{4\pi^2}\frac{k_{\bot}^2}{\upsilon}
\bigg[(1-e^{-2\upsilon{D}})|\gamma^+|^2+(e^{2\upsilon{D}}-1)|\gamma^-|^2)\bigg]
\ ,
\end{eqnarray}

\begin{eqnarray}
\nonumber
E_{vac}=\frac{1}{8\pi}\int\frac{d^2k_{\bot}}{4\pi^2}k_{\bot}
\bigg(e^{-2k_{\bot}{D}}|\phi^-|^2+|\phi^+|^2\bigg) \ .
\end{eqnarray}

Here $\alpha=x,y,z$ and
\begin{eqnarray}
\nonumber G_x({\bf k})=G_y({\bf
k})=\frac{1}{1+k_{\bot}^2\lambda_c^2+k_z^2\lambda_{ab}^2} \ , \
G_z({\bf
k})=\frac{1+k^2\lambda_c^2}{(1+k^2\lambda_{ab}^2)(1+k_{\bot}^2\lambda_c^2+k_z^2\lambda_{ab}^2)}
\ ,
\end{eqnarray}

\begin{eqnarray}
\nonumber \gamma^-({\bf
k}_{\bot})=\frac{\upsilon\bigg[A(k_{\bot}-\upsilon)e^{-\upsilon
D}-B(k_{\bot}+\upsilon)\bigg]}{k_{\bot}C} \ , \  \gamma^+({\bf
k}_{\bot})=\frac{\upsilon\bigg[A(k_{\bot}+\upsilon)e^{\upsilon
D}-B(k_{\bot}-\upsilon)\bigg]}{k_{\bot}C} \ ,
\end{eqnarray}

\begin{eqnarray}
\nonumber \phi^-({\bf k}_{\bot})=(\upsilon/k_{\bot}C)
\times\bigg[-2k_{\bot}B+\bigg[(k_{\bot}+\upsilon)e^{\upsilon
D}+(k_{\bot}-\upsilon)e^{-\upsilon D}\bigg]A\bigg] \ ,
\\
\nonumber \phi^+({\bf k}_{\bot})=(-\upsilon/k_{\bot}C)
\bigg[-2k_{\bot}A+\bigg[(k_{\bot}+\upsilon)e^{\upsilon
D}+(k_{\bot}-\upsilon)e^{-\upsilon D}\bigg]B\bigg] \ ,
\end{eqnarray}

\begin{eqnarray}
\nonumber C(k_{\bot})=e^{-\upsilon
D}(k_{\bot}-\upsilon)^2-e^{\upsilon D}(k_{\bot}+\upsilon)^2 \ , \
\upsilon=\sqrt{k_{\bot}^2+\lambda_{ab}^{-2}} \ ,
\end{eqnarray}

\begin{eqnarray}
\nonumber A({\bf k}_{\bot})=\frac{1}{2D}\sum_mg_l({\bf
k}_{\bot},k_m)\nu_z^{vi}({\bf k}_{\bot},k_m) \ , \  B({\bf
k}_{\bot})=\frac{1}{2D}\sum_me^{-ik_md}g_l({\bf
k}_{\bot},k_m)\nu_z^{vi}({\bf k}_{\bot},k_m) \ ,
\end{eqnarray}

\begin{eqnarray}
\nonumber g_l(k)=\frac{\phi_0}{1+k^2\lambda_{ab}^2} \ .
\end{eqnarray}

The summation in the above expressions is carried out over
$k_z=k_m\equiv m\pi/D,\,\,m=0,\pm 1,\pm 2, ...$ and $\bot$ stands
for the vector component parallel to the $xy$ plane. Following
Ref.~\onlinecite{Brandt-prb00} we introduce here the Fourier transform
${\bf \nu}^{vi}({\bf k}_{\bot},k_m)$ of the vorticity distribution
${\bf\nu}^{vi}(\bf r)$:

\begin{eqnarray}
{\bf \nu_{\bot}}^{vi}({\bf k}_{\bot},k_m)=-2i\int
d^2r_{\bot}e^{-i{\bf k_{\bot}r_{\bot}}}\int_{-D}^0 dz
\sin(k_mz){\bf \nu}_{\bot}^{vi}({\bf r}_{\bot},z) \ ,
\label{eq:b2}
\\
{ \nu_{z}}^{vi}({\bf k}_{\bot},k_m)=2\int d^2r_{\bot}e^{-i{\bf
k_{\bot}r_{\bot}}}\int_{-D}^0 dz \cos(k_mz){ \nu}_z^{vi}({\bf
r}_{\bot},z) \ . \label{eq:b3}
\end{eqnarray}
For a pair of curved vortices shifted in the $y$ direction at a
certain distance $R$ the expressions \eqref{eq:b2} and
\eqref{eq:b3} take the form:

\begin{eqnarray}
\nonumber \nu_{x}^{vi}({\bf k}_{\bot},k_m)=0 \ ,
\\
\nonumber \nu_{y}^{vi}({\bf
k}_{\bot},k_m)=-2i\bigg(1+e^{-ik_yR}\bigg)\int_{-D}^{0}y'(z)e^{-ik_yy(z)}\sin(k_mz)dz
\ ,
\\
\nonumber \nu_{z}^{vi}({\bf
k}_{\bot},k_m)=2\bigg(1+e^{-ik_yR}\bigg)\int_{-D}^{0}e^{-ik_yy(z)}\cos(k_mz)dz
\ .
\end{eqnarray}
To find the vortex--vortex interaction energy we should take the
terms in Eq.~\eqref{eq:b0} which depend on mutual vortex
arrangement:

\begin{equation}\label{eq:b6}
\varepsilon_{int}=\varepsilon_0
\left(\varepsilon_{int}^{vi}+\varepsilon_{int}^{stray}+\varepsilon_{int}^{vac}\right)\,,
\end{equation}
where $\varepsilon_0=\phi_0^2/16\pi^3\lambda_{ab}$ and

\begin{eqnarray}\label{eq:b7}
\epsilon_{int}^{vi}&=&\pi\int_{0}^{\infty} \nu d\nu
    \int_{-\tilde D}^{0} d\zeta_1 \int_{-\tilde D}^{0} d\zeta_2\,
    \bigg\{
        J_0\left[\nu\left(\eta(\zeta_2)-\eta(\zeta_1) +
            \tilde R \right)\right] +
        J_0\left[\nu\left(\eta(\zeta_2)-\eta(\zeta_1) -
            \tilde R \right)\right]\bigg\} \nonumber \\ &&\qquad\times\left(\Pi_1(\nu,\zeta_1,\zeta_2)
                +\Pi_2(\nu,\zeta_1,\zeta_2)\right)\ ,
\end{eqnarray}
\begin{eqnarray}\label{eq:b8}
\epsilon_{int}^{stray}&=&4\pi\int_{0}^{\infty} \nu d\nu %
    \int_{-\tilde D}^{0} d\zeta_1 \int_{-\tilde D}^{0} d\zeta_2\,
    \bigg\{
        J_0\left[\nu\left(\eta(\zeta_2)-\eta(\zeta_1) +
            \tilde R \right)\right] +
        J_0\left[\nu\left(\eta(\zeta_2)-\eta(\zeta_1) -
            \tilde R \right)\right]\bigg\} \nonumber \\
    &&\qquad\times\frac{\Pi_3(\nu,\zeta_1,\zeta_2)}
{\tau(\nu)\sinh\left[\tau(\nu)\tilde
D\right]\bigg\{e^{-\tau(\nu)\tilde D}\left[\nu-\tau(\nu)\right]^2-
e^{\tau(\nu)\tilde D}\left[\nu+\tau(\nu)\right]^2\bigg\}^2} \ ,
\end{eqnarray}
\begin{eqnarray}\label{eq:b9}
\epsilon_{int}^{vac}&=&4\pi\int_{0}^{\infty} \nu d\nu
    \int_{-\tilde D}^{0} d\zeta_1 \int_{-\tilde D}^{0} d\zeta_2\,
    \bigg\{
        J_0\left[\nu\left(\eta(\zeta_2)-\eta(\zeta_1) +
            \tilde R \right)\right] +
        J_0\left[\nu\left(\eta(\zeta_2)-\eta(\zeta_1) -
            \tilde R \right)\right]\bigg\} \nonumber\\
    &&\qquad\times\frac{\Pi_4(\nu,\zeta_1,\zeta_2)}
{\nu\sinh^2\left[\tau(\nu)\tilde
D\right]\bigg\{e^{-\tau(\nu)\tilde D}\left[\nu-\tau(\nu)\right]^2-
        e^{\tau(\nu)\tilde D}\left[\nu+\tau(\nu)\right]^2\bigg\}^2} \,,
\end{eqnarray}
$$
    \tau(\nu)=\sqrt{1+\Gamma^2\nu^2}\,,\qquad u(\nu)=\sqrt{1+\nu^2}\,.
$$
Here we have introduced dimensionless coordinates $\eta = y
/\lambda_{ab}$, $\zeta = z / \lambda_{ab}$,  dimensionless wave
number $\nu = q \lambda_{ab}$, and use the notations

\begin{eqnarray} \nonumber
\Pi_1(\nu,\zeta_1,\zeta_2) = \eta'(\zeta_1) \eta'(\zeta_2)
\frac{\sinh\left[\tau(\nu)\left(\tilde
D+(\zeta_1+\zeta_2-|\zeta_1-\zeta_2|)/2\right)\right]
\sinh\left[\tau(\nu)\left((-\zeta_1-\zeta_2-|\zeta_1-\zeta_2|)/2\right)\right]}
    {\tau(\nu)\sinh\left[\tilde D\tau(\nu)\right]} \nonumber \ ,
\end{eqnarray}

\begin{eqnarray}\nonumber
\Pi_2(\nu,\zeta_1,\zeta_2)&=&\frac{\tau (\nu)
\cosh\left[\tau(\nu)\left(\tilde
D+(\zeta_1+\zeta_2-|\zeta_1-\zeta_2|)/2\right)\right]\,
\cosh\left[\tau(\nu)\left((-\zeta_1-\zeta_2-|\zeta_1-\zeta_2|)/2\right)\right]}
    {\nu^2\sinh\left[\tilde D\tau (\nu)\right]} \nonumber \\
&-&\frac{\cosh\left[u(\nu)\left(\tilde
D+(\zeta_1+\zeta_2-|\zeta_1-\zeta_2|)/2\right)\right]
\cosh\left[u(\nu)\left((-\zeta_1-\zeta_2-|\zeta_1-\zeta_2|)/2\right)\right]}
{\nu^2u(\nu)\sinh\left[\tilde Du (\nu)\right]} \nonumber \ ,
\end{eqnarray}

\begin{eqnarray}\nonumber
\Pi_3(\nu,\zeta_1,\zeta_2)&=&\bigg\{\nu^2\cosh\left[\tau(\nu)
    \tilde D\right]+\tau (\nu)^2\cosh\left[\tau (\nu)\tilde D\right] +
    2\nu\tau(\nu)\sinh\left[\tau (\nu)\tilde D\right]\bigg\} \nonumber\\
&&\times\bigg\{\cosh\left[\tau(\nu)(\tilde
D+\zeta_1)\right]\cosh\left[\tau(\tilde D+\zeta_2)\right]
+\cosh\left[\tau(\nu)\zeta_1\right]\cosh\left[\tau(\nu)\zeta_1\right]\bigg\}\nonumber\\
&+&\cosh\left[\tau(\nu)(\tilde
D+\zeta_1)]\right)\cosh\left[\tau(\nu)\zeta_2\right] +
\cosh\left[\tau(\nu)z_1\right]\cosh\left[\tau(\nu)(\zeta_2+\tilde
D)\right]\nonumber\,
\end{eqnarray}
\begin{eqnarray}
\nonumber \Pi_4(\nu,\zeta_1,\zeta_2)=\nonumber
\bigg\{2\nu^2\cosh^2\left[\tau(\nu)\tilde
D\right]+\sinh^2\left[\tau(\nu)\tilde D\right]+
2\nu\tau(\nu)\sinh\left[\tau(\nu)\tilde
D\right]\cosh\left[\tau(\nu)\tilde D\right]\bigg\}\times
\\
\nonumber \bigg\{\cosh\left[\tau(\nu)(\tilde
D+\zeta_1)\right]\cosh\left[\tau(\nu)(\tilde D+\zeta_2)\right]+
\cosh\left[\tau(\nu)\zeta_1\right]\cosh\left[\tau(\nu)\zeta_1\right]\bigg\}-
\\
\nonumber 2\bigg\{\nu^2\cosh\left[\tau(\nu)\tilde
D\right]+\nu\tau\sinh\left[\tau(\nu)\tilde D\right]\bigg\}\times
 \\
 \nonumber
\bigg\{\cosh\left[\tau(\nu)(\tilde
D+\zeta_1)\right]\cosh\left[\tau(\nu)\zeta_2\right]+
\cosh\left[\tau(\nu)\zeta_1\right]\cosh\left[\tau(\nu)(\zeta_2+\tilde
D)\right]\bigg\} \ .
\end{eqnarray}
The dimensionless thickness of the film $\tilde D$ and
dimensionless intervortex distance
 $\tilde R$ are measured in the units of $\lambda_{ab}$.

\newpage
%
\begin{figure}[t!]
\includegraphics[width=0.45\textwidth]{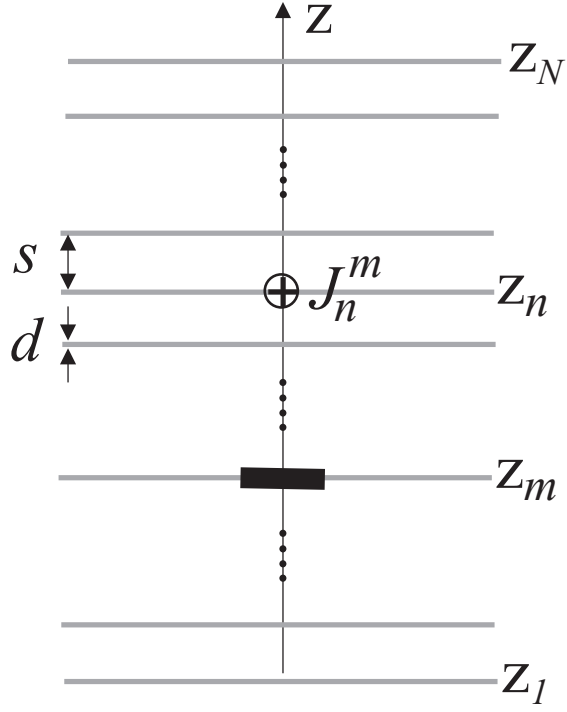}
\caption{A single $2D$ pancake vortex positioned in the $m-$th
layer of a finite layered structure, $d$ is a thickness of the
superconducting layer, and $s$ is the distance between the
layers.} \label{Fig:1}
\end{figure}
%

%
\begin{figure}
\includegraphics[width=0.40\textwidth]{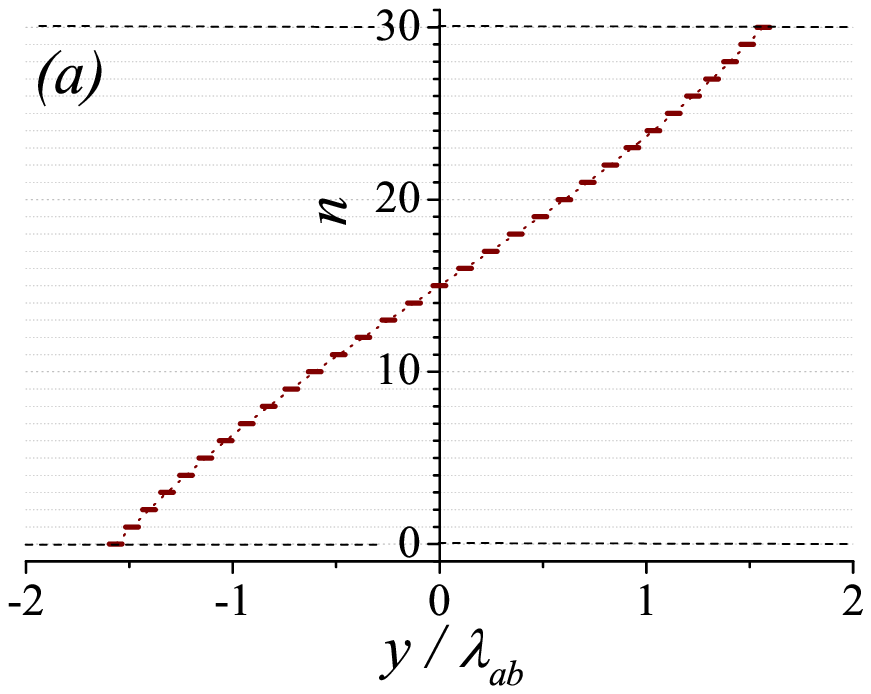}
\includegraphics[width=0.40\textwidth]{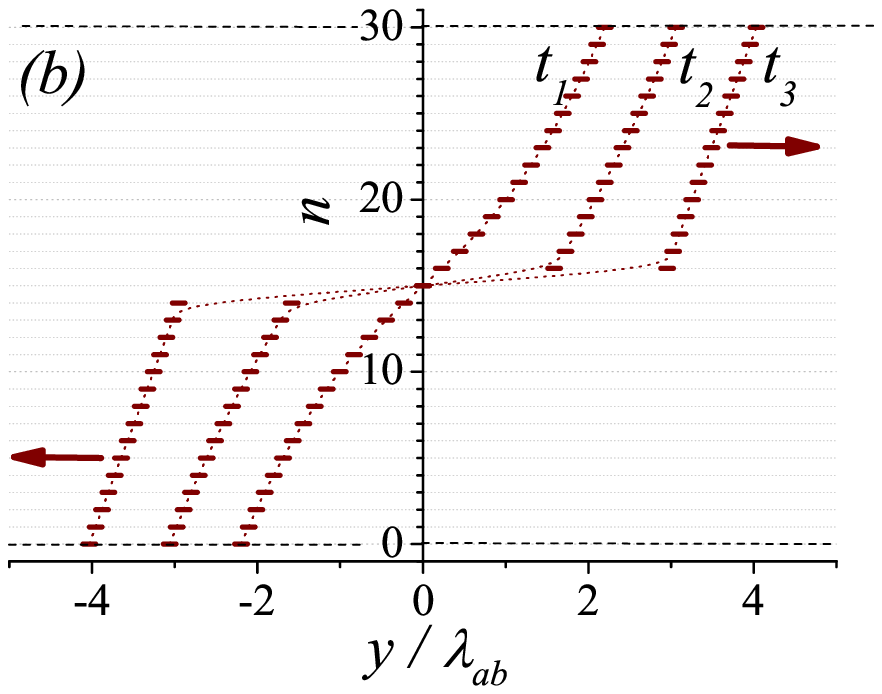}
\includegraphics[width=0.40\textwidth]{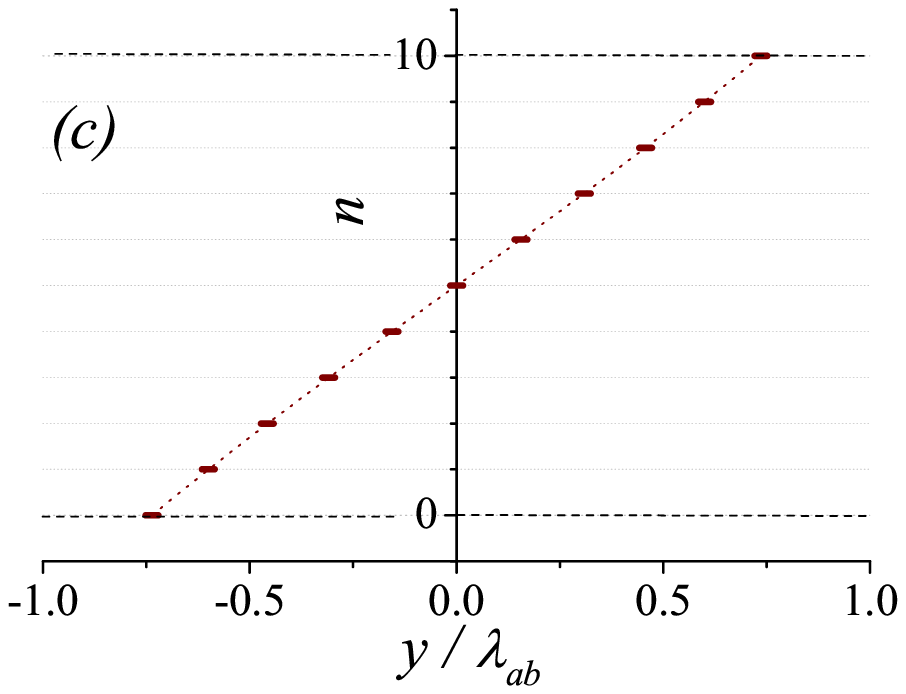}
\includegraphics[width=0.37\textwidth]{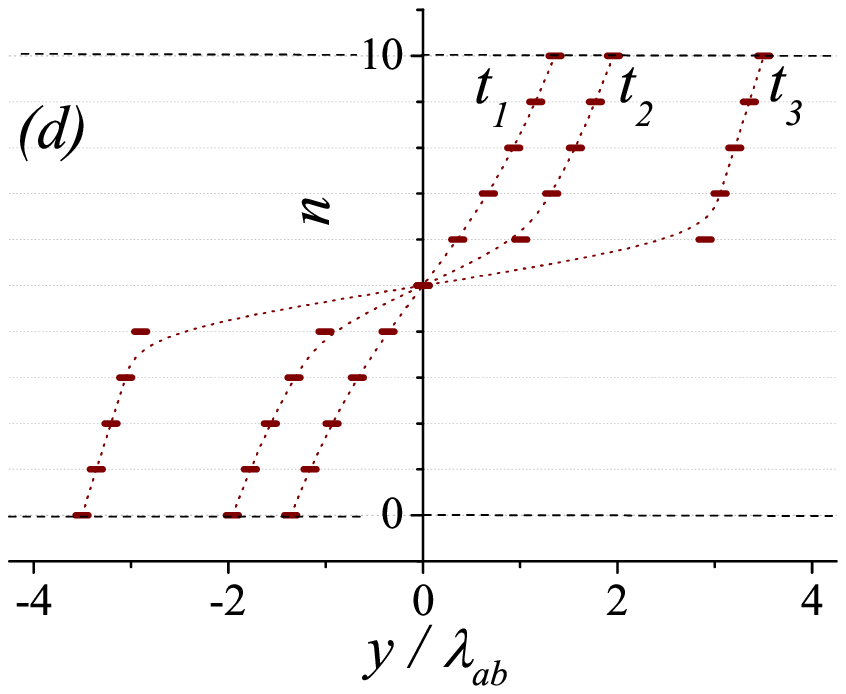}
\caption{(Color online) Configurations of $N=31$ (panels a, b) and
$N=11$ (panels c, d) pancakes in a finite stack in the presence of
the applied in-plane magnetic field $H_a$. (a) The force-balanced
(equilibrium) configuration of pancakes for $H_a = 0.2 H_0 < H^*$.
(b) Pancake configurations at sequential time points $t_1 < t_2 <
t_3$ for $H_a = 0.22 H_0 > H^*$. For the structure with $N=31$ we
find $H^* \simeq 0.21 H_0$.
  (c) The force-balanced (equilibrium) configuration of
pancakes for $H_a = 0.35 H_0 < H^*$. (d) Pancake configurations at
sequential time points $t_1 < t_2 < t_3$ for $H_a = 0.4 H_0 >
H^*$. For the structure with $N=11$ we find $H^* \simeq 0.38 H_0$.
Here $H_0 = \phi_0 / 2\pi \lambda_{ab}^2$, $\Lambda = 10
\lambda_{ab}$, and $s = 0.1 \lambda_{ab}$.} \label{Fig:2}
\end{figure}

%
\begin{figure}[h]
\includegraphics[width=0.45\textwidth]{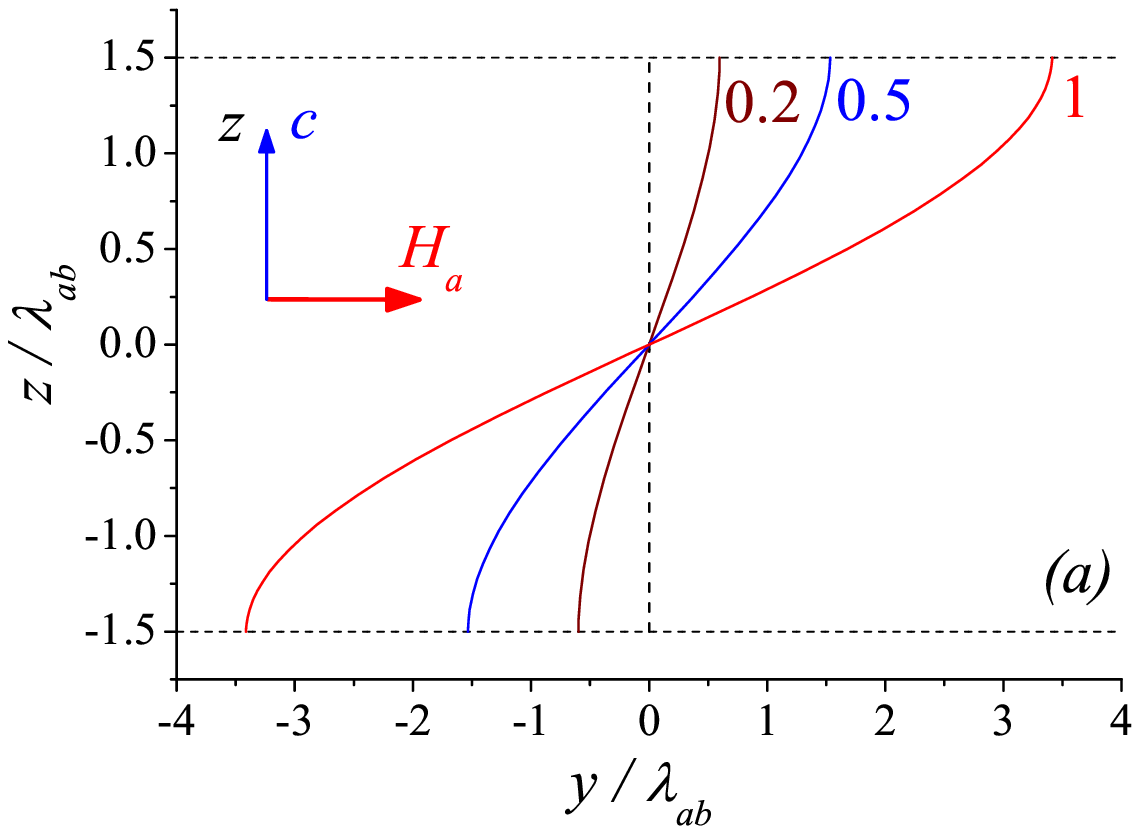}
\includegraphics[width=0.45\textwidth]{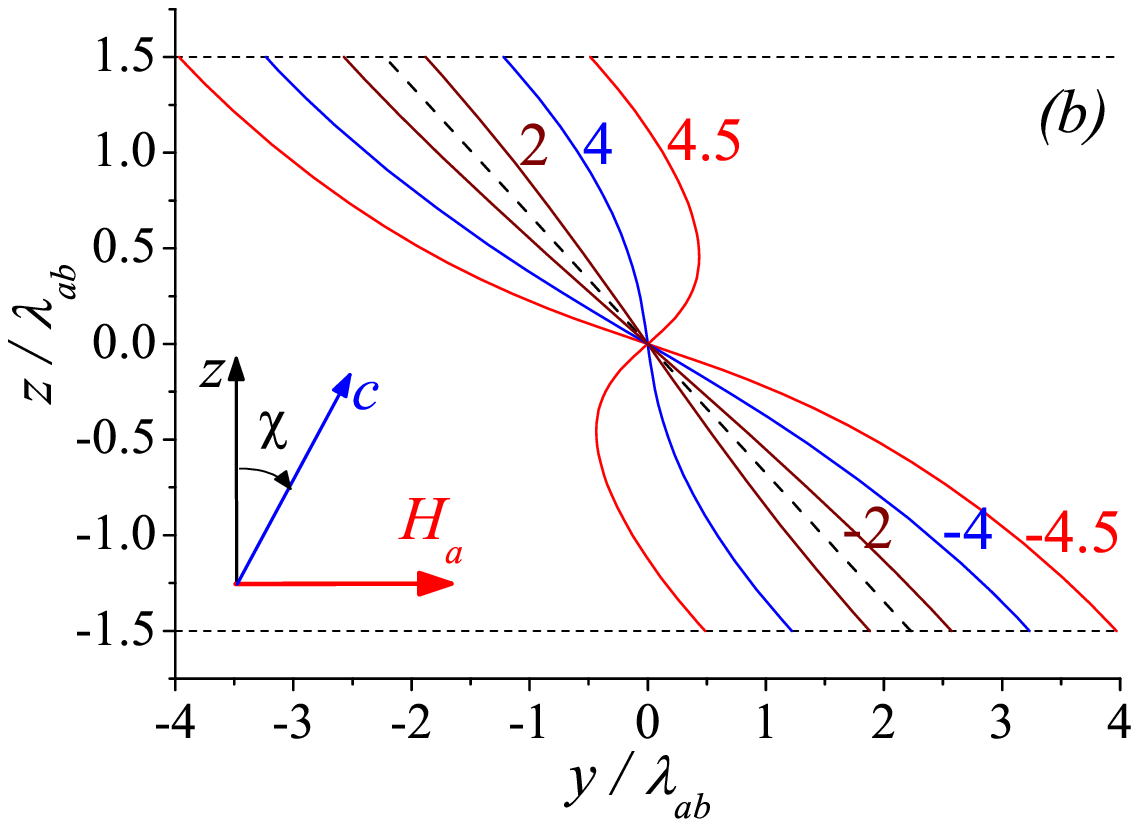}
\caption{(Color online) Typical configurations of the vortex lines in the film of
the thickness $D=3\lambda_{ab}$ for the anisotropy parameter
$\Gamma=5$ and for different values of in-plane magnetic field
$H_\| = H_a\,\mathbf{y}_0$. (a) The anisotropy axis is
perpendicular to the film plane ($\chi=0^o$, $H_{c1}^{(0)} \simeq
1.74\, H_{ab} $). (b) The anisotropy axis is tilted with respect
to the $z$ axis ($\chi=30^o$, $H_{c1}^{(\chi)} \simeq 4.6\, H_{ab}
$) The numbers near the curves denote the values of the ratio $H_a
/ H_{ab}$. The dashed line shows the shape of a vortex line in the
absence of the in-plane magnetic field.} \label{Fig:3}
\end{figure}
%

%
\begin{figure}
\includegraphics[width=0.5\textwidth]{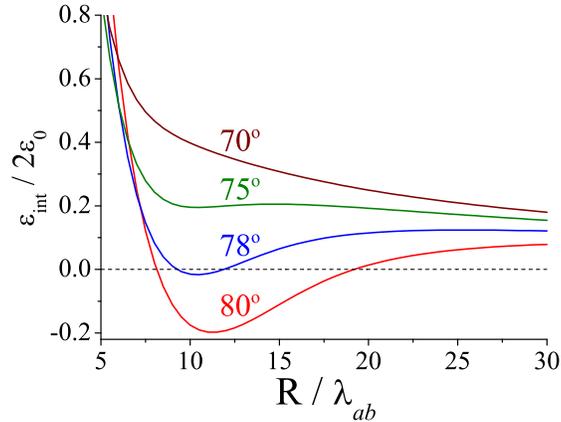}
\caption{(Color online) Typical plots of the interaction energy
per vortex [Eqs.~(\ref{eq:10c}) and(\ref{eq:11c})] vs the distance R between two vortices for a film of
thickness $d=3\lambda_{ab}$ and different tilting angles
$\gamma=70^o,\, 75^o,\, 78^o,\, 80^o$
$(\varepsilon_0=\phi_0^2/16\pi^3\lambda_{ab})$ .} \label{Fig:4}
\end{figure}
%
%
\newpage
%
\begin{figure}[h]
\includegraphics[width=0.4\textwidth]{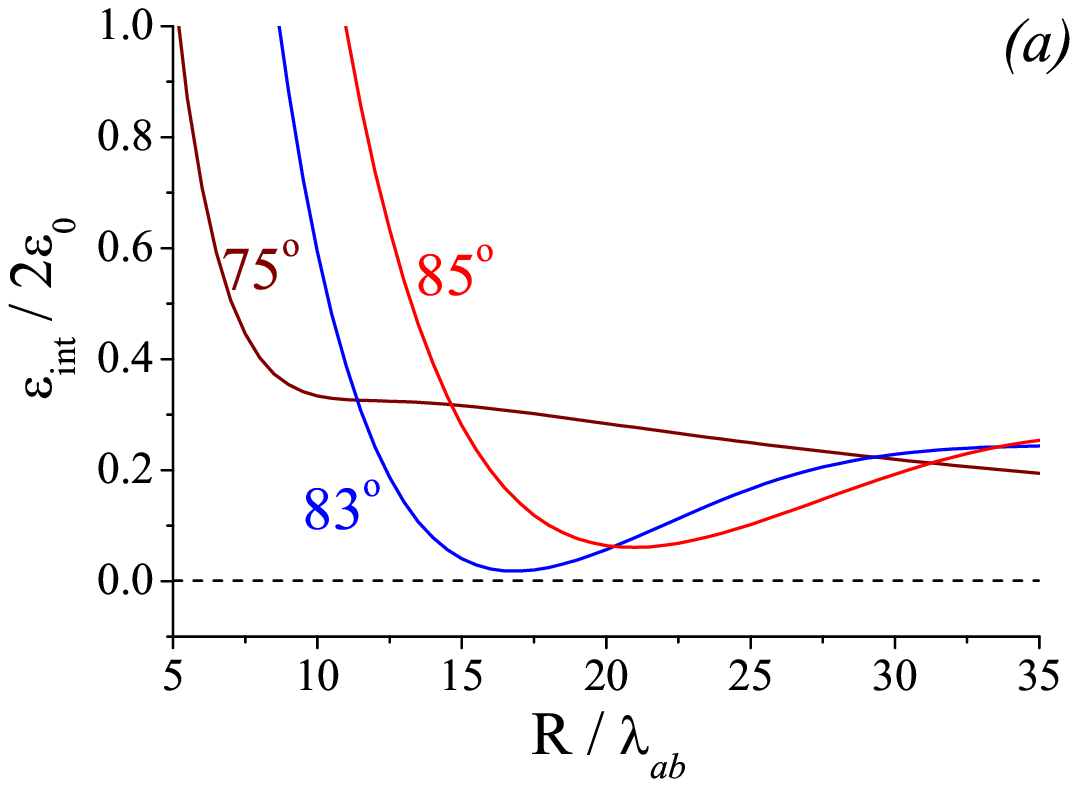}
\includegraphics[width=0.4\textwidth]{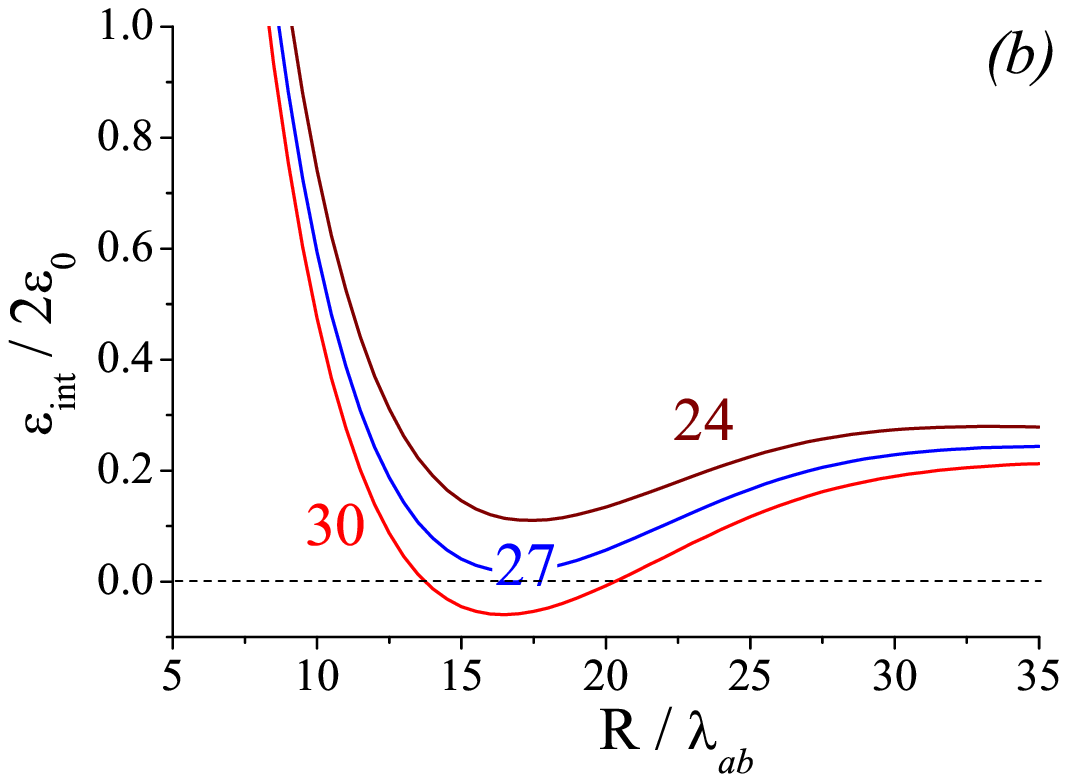}
\caption{(Color online) Typical plots of the interaction energy
per vortex [Eqs.~(\ref{straight-tilted}),(\ref{straight-tilted2})]
vs the distance R between two tilted vortices for an anisotropic
film of the thickness $D=3\lambda_{ab}$. (a) Interaction energy
for the anisotropy parameter $\Gamma=27$ and different tilting
angles. The numbers near the curves denote the values of tilting
angle $\gamma$. (b) Interaction energy for $\gamma=83^o$ and
different values of anisotropy parameter. The numbers near the
curves denote the values of $\Gamma$.} \label{Fig:5}
\end{figure}
%
%
\begin{figure}[h]
\includegraphics[width=0.4\textwidth]{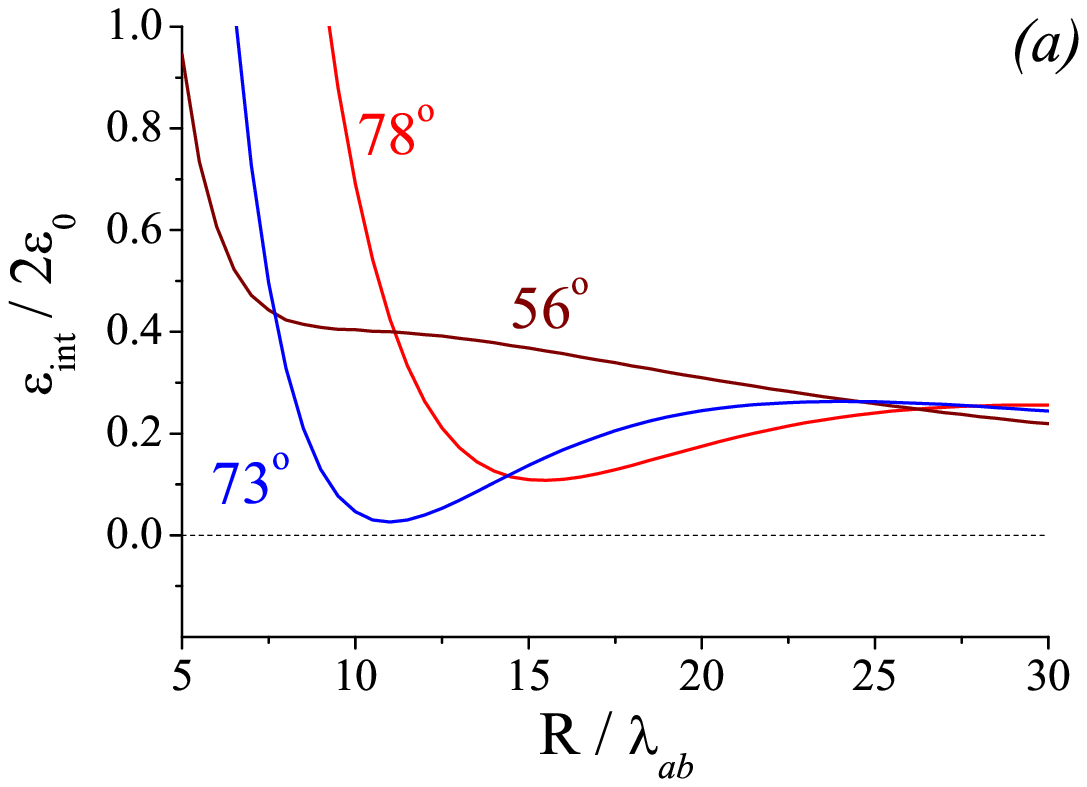}
\includegraphics[width=0.4\textwidth]{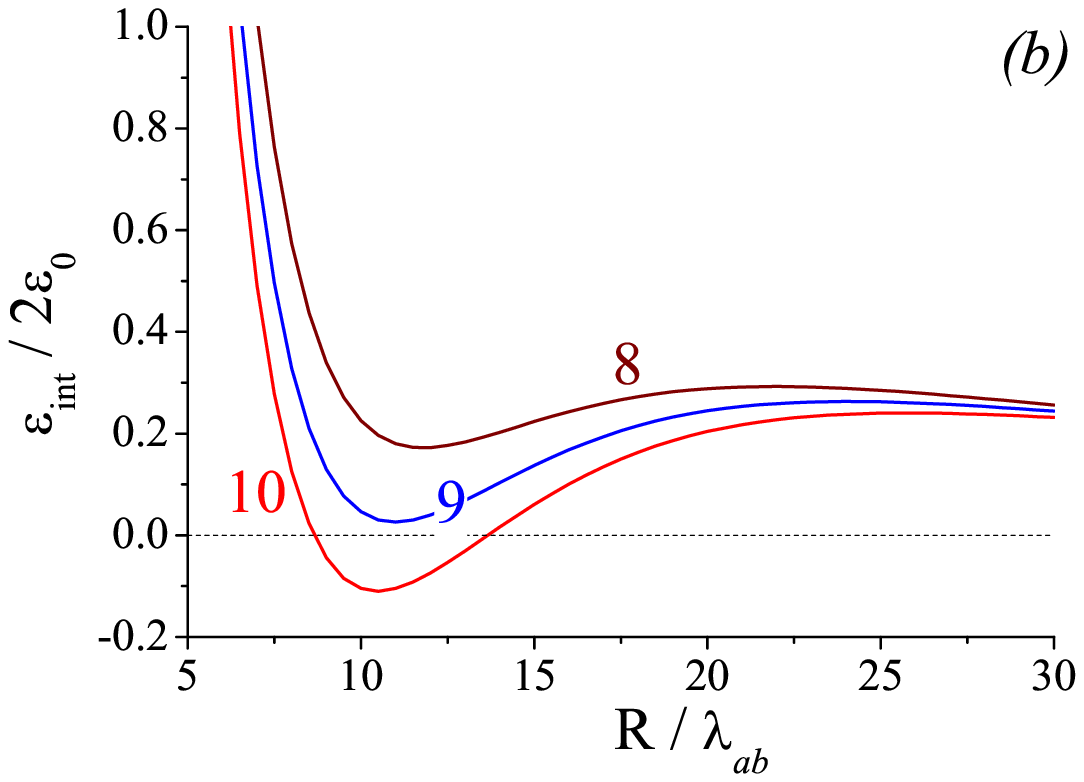}
\caption{(Color online) Typical plots of the interaction energy
per vortex [Eqs.~(\ref{straight-tilted}),(\ref{straight-tilted2})]
vs the distance R between two tilted vortices for an anisotropic
film of the thickness $D=10\lambda_{ab}$. (a) Interaction energy
for the anisotropy parameter $\Gamma=9$ and different tilting
angles. The numbers near the curves denote the values of tilting
angle $\gamma$. (b) Interaction energy for $\gamma=73^o$ and
different values of anisotropy parameter. The numbers near the
curves denote the values of $\Gamma$.} \label{Fig:51}
\end{figure}
%
\pagebreak
\begin{figure}[b]
\includegraphics[width=0.45\textwidth]{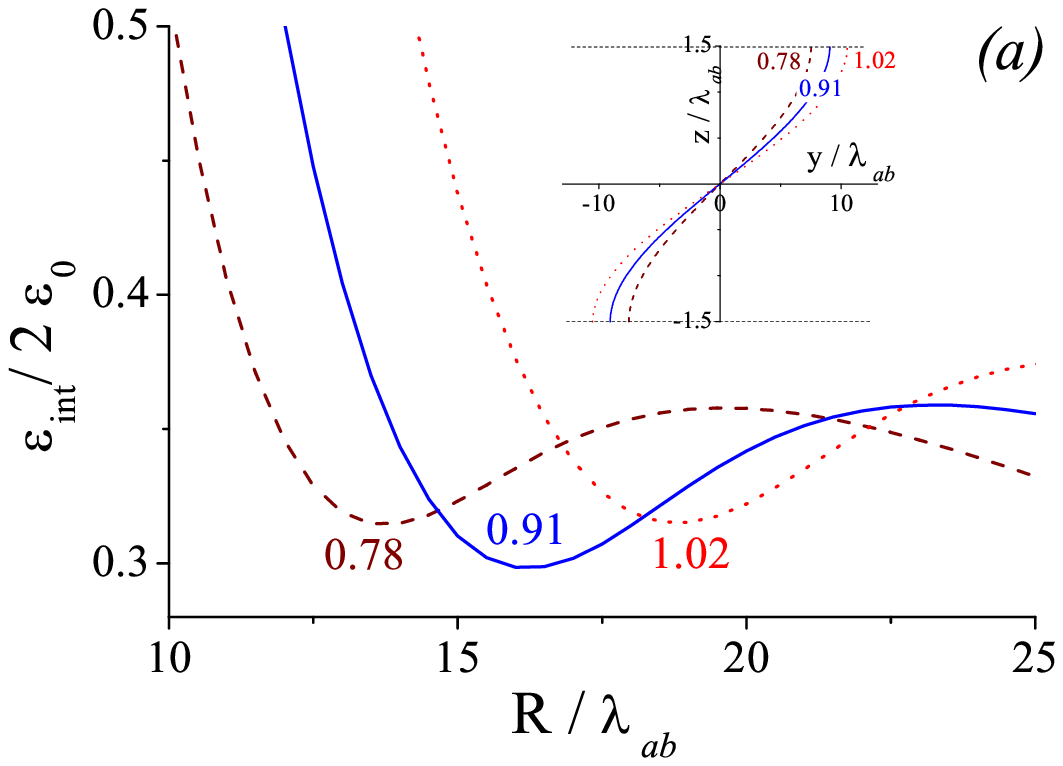}
\includegraphics[width=0.45\textwidth]{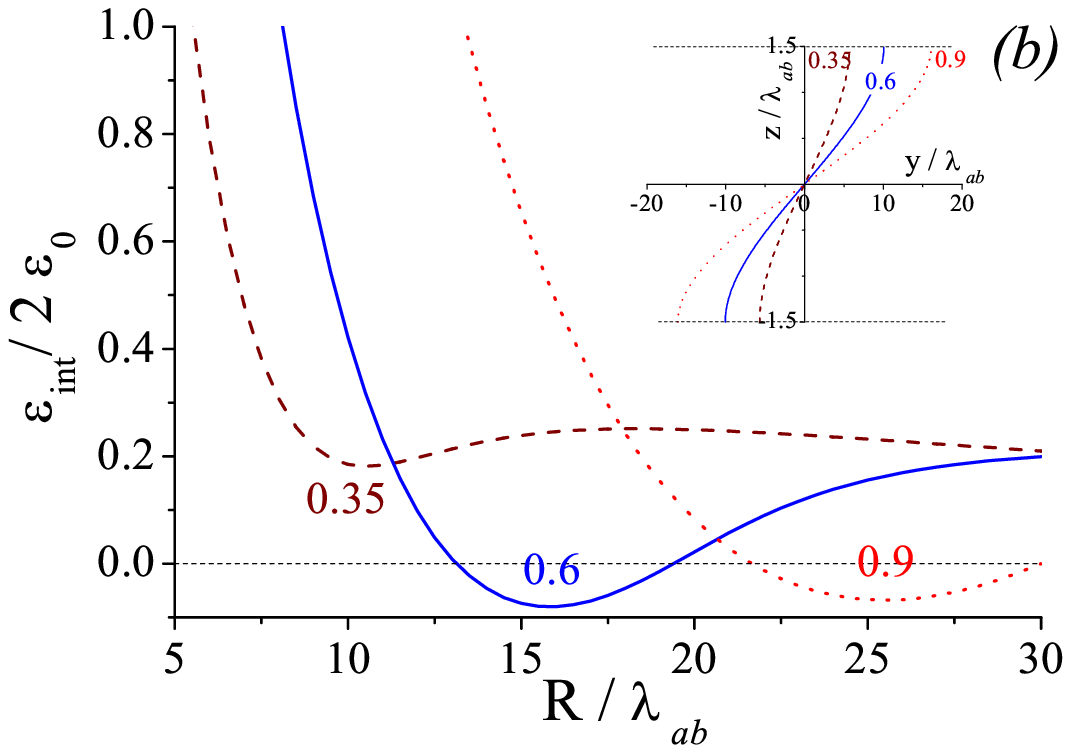}
\caption{(Color online) Typical plots of the interaction energy
per vortex (\ref{int energy}) vs the distance $R$ between two
curved vortices for an anisotropic film of the thickness
$D=3\lambda_{ab}$: (a) $\Gamma=15$ ; (b) $\Gamma=27$. The numbers
near the curves denote the values of the ratio $H_a/H_{ab}$. The
shape of vortex lines is schematically shown in the insets.}
\label{Fig:6}
\end{figure}
%
\begin{figure}[b]
\includegraphics[width=0.45\textwidth]{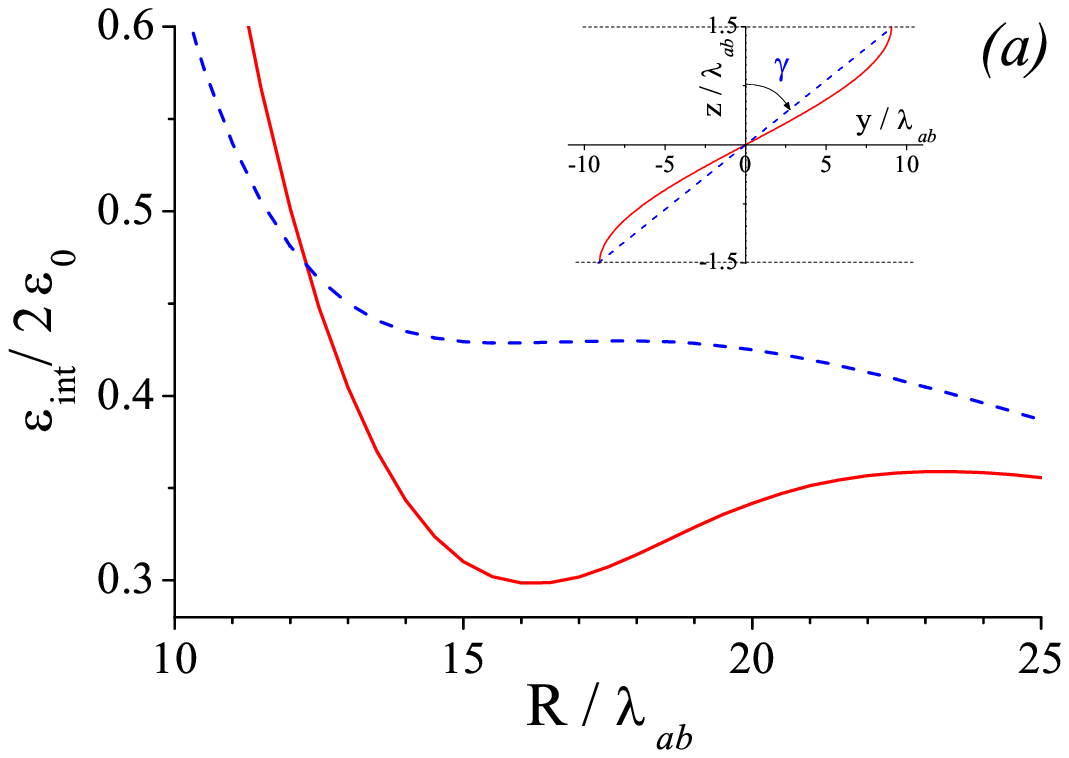}
\includegraphics[width=0.45\textwidth]{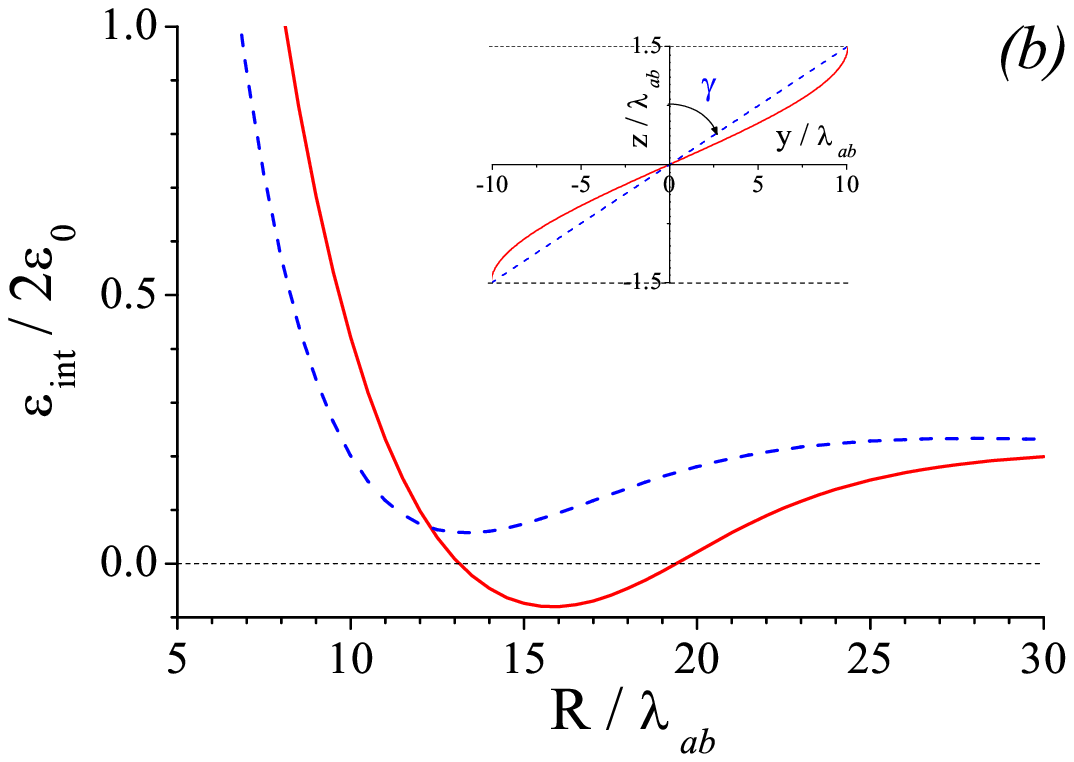}
\caption{(Color online) Comparison of the vortex--vortex
interaction potentials for curved [Eq.~(\ref{int energy})] (solid
lines) and straight tilted [Eqs.~(\ref{straight-tilted}),
(\ref{straight-tilted2})] (dashed lines) vortices  for an
anisotropic film of the thickness $D=3\lambda_{ab}$ with different
anisotropy parameters: (a) $\Gamma=15$, $H_a=0.91 H_{ab}$
($\gamma=80.6^o)$; (b) $\Gamma=27$, $H_a=0.6 H_{ab}$
($\gamma=81.5^o)$. The shape of vortex lines is schematically
shown in the insets.} \label{Fig:7}
\end{figure}
%

%
\begin{figure}
\includegraphics[width=0.45\textwidth]{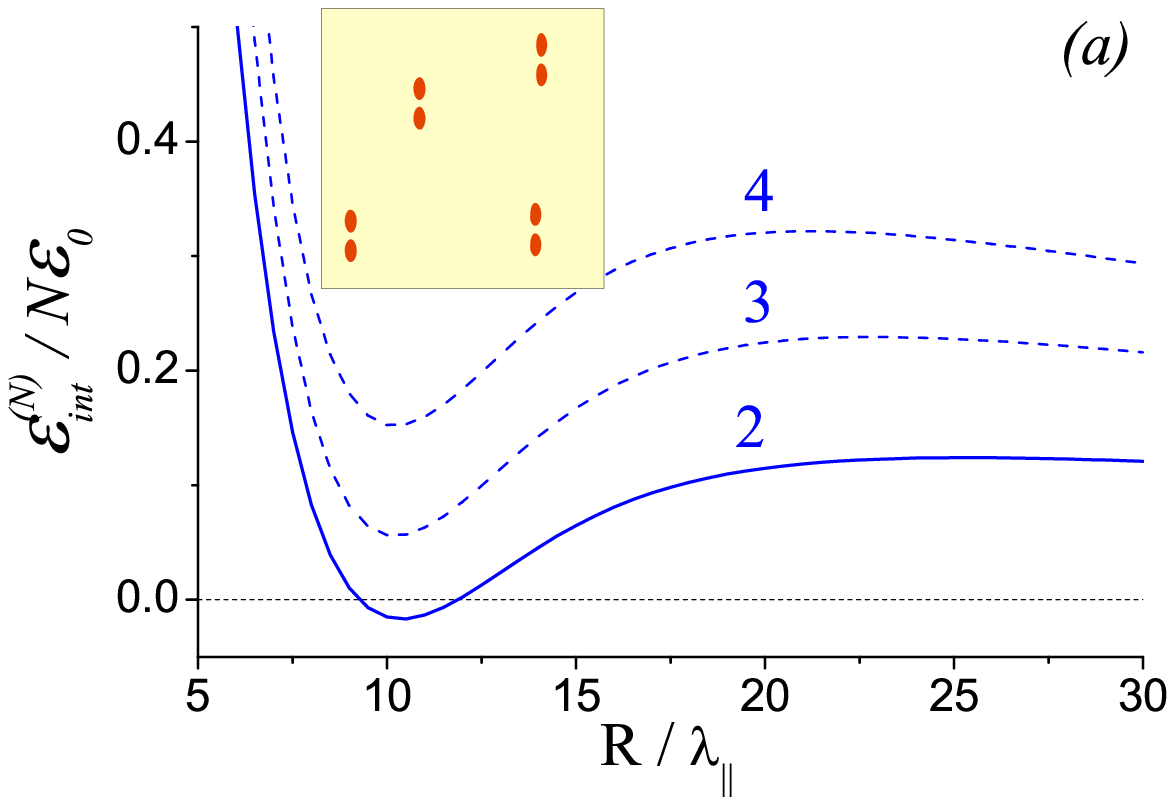}
\includegraphics[width=0.45\textwidth]{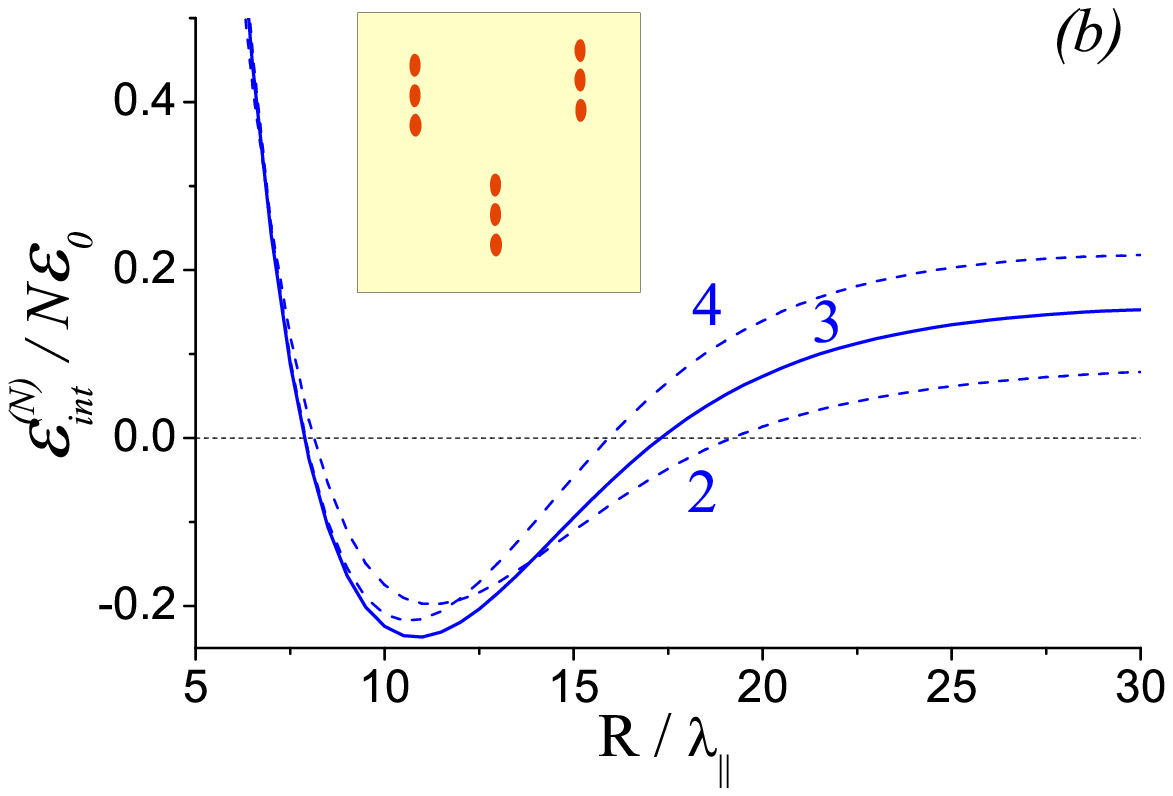}
\caption{(Color online) Typical plots of the interaction energy
per vortex [Eqs.~(\ref{eq:10c}),(\ref{eq:11c}),(\ref{eq:1d})]
 vs the intervortex distance $R$ in an equidistant chain of N vortices in a
 stack of decoupled superconducting layers ($D=3\lambda_{ab}$): (a) $\gamma=78^o$; (b) $\gamma=80^o$.
The numbers near the curves denote the number $N$ of vortices in
molecule. Inserts show schematic pictures of vortex matter
consisting of dimeric (a) and trimeric (b) molecules.}
\label{Fig:8}
\end{figure}
%
%
\begin{figure}
\includegraphics[width=0.5\textwidth]{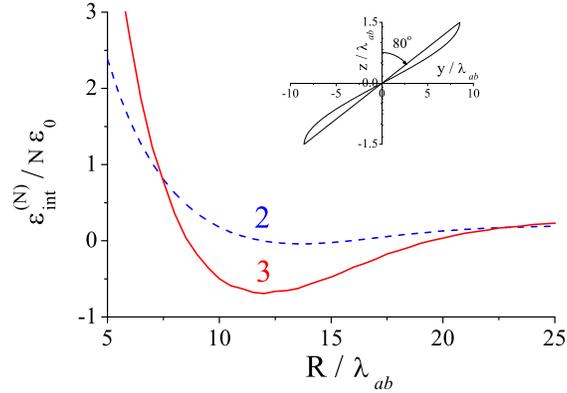}
\caption{(Color online) Typical plots of the interaction energy
per vortex vs the intervortex distance R in an equidistant chain
of N vortices for $d=3\lambda_{ab}$, $\Gamma=27$, $H_a=0.513H_{ab}$($\gamma=80^o$).
The numbers near the curves denote the number N of vortices in
a molecule. The shape of vortex line and effective tilting angle $\gamma$
 are schematically shown in the inset.} \label{Fig:9}
\end{figure}
%

\begin{figure}
\includegraphics[width=0.3\textwidth]{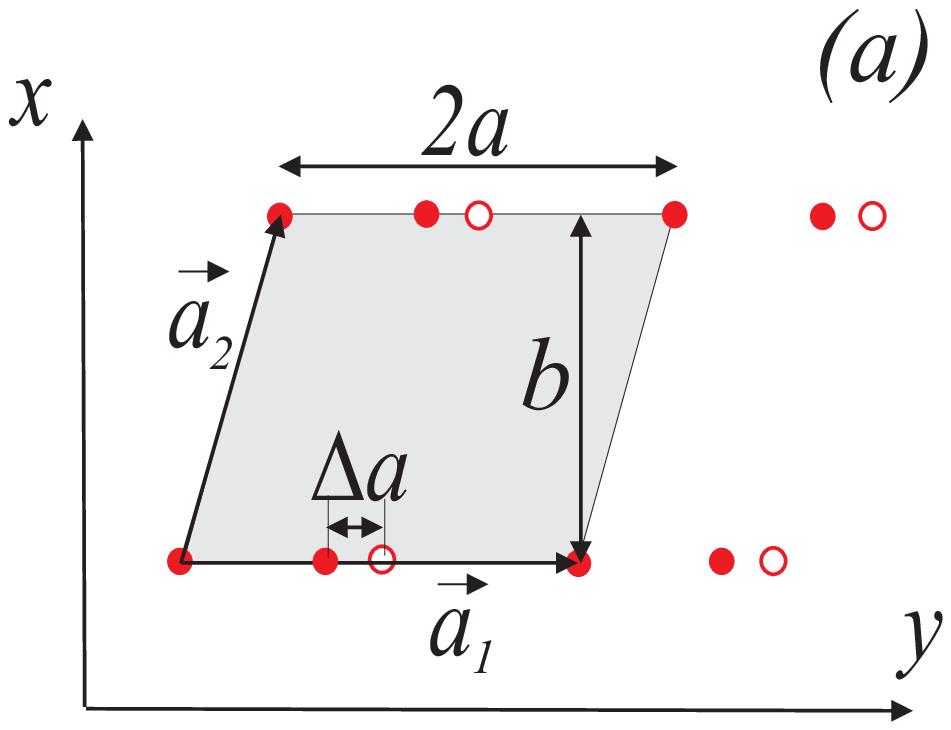}
\includegraphics[width=0.37\textwidth]{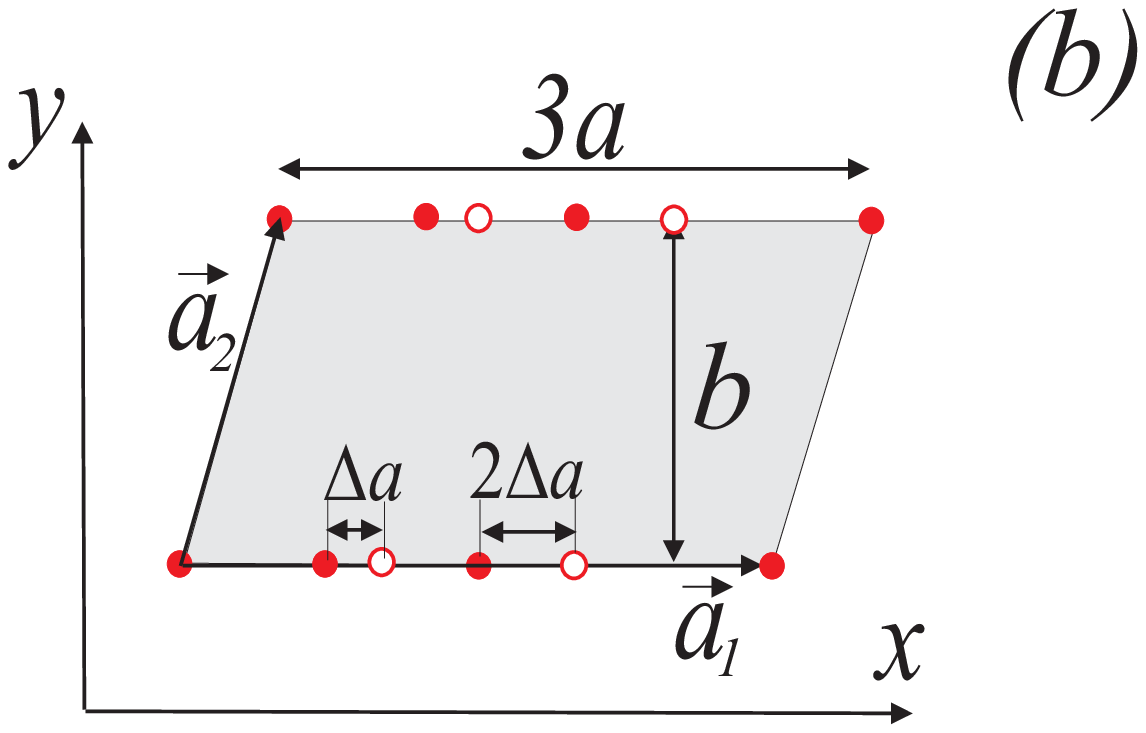}
\caption{(Color online) Vortex lattice with two $M=2$ ({\it a}) and three $M=3$ ({\it b})
vortices per a primitive cell.} \label{Fig:10}
\end{figure}

%
\begin{figure}[t]
\includegraphics[width=0.45\textwidth]{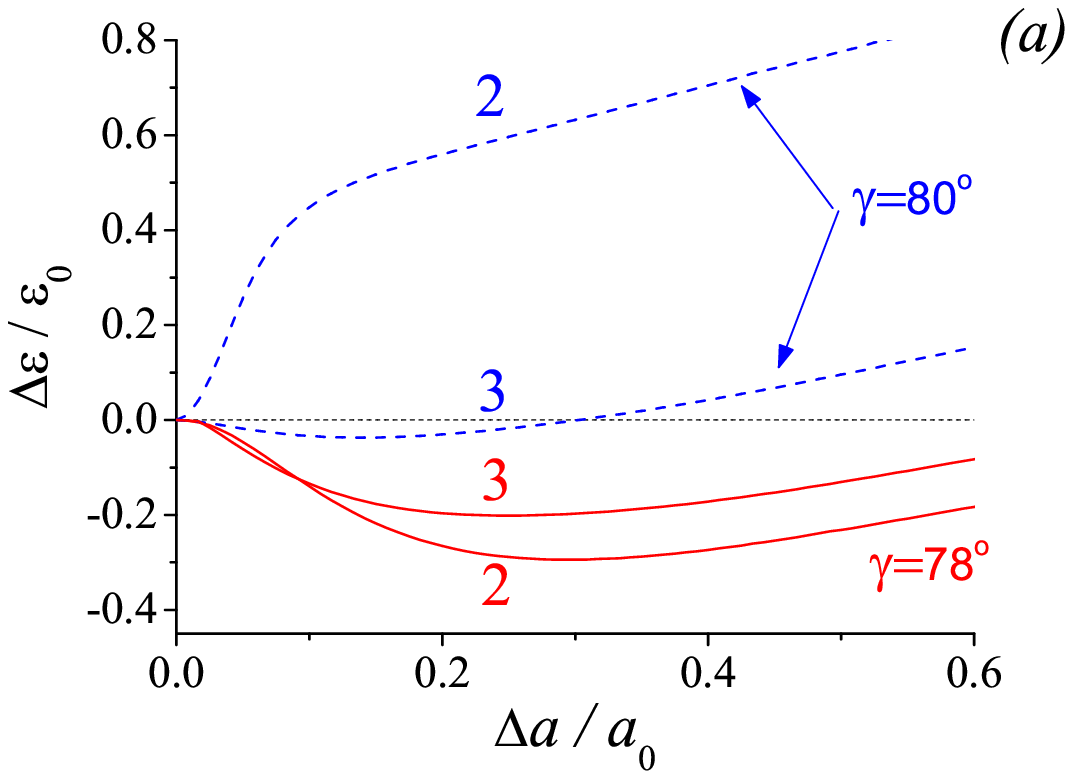}
\includegraphics[width=0.45\textwidth]{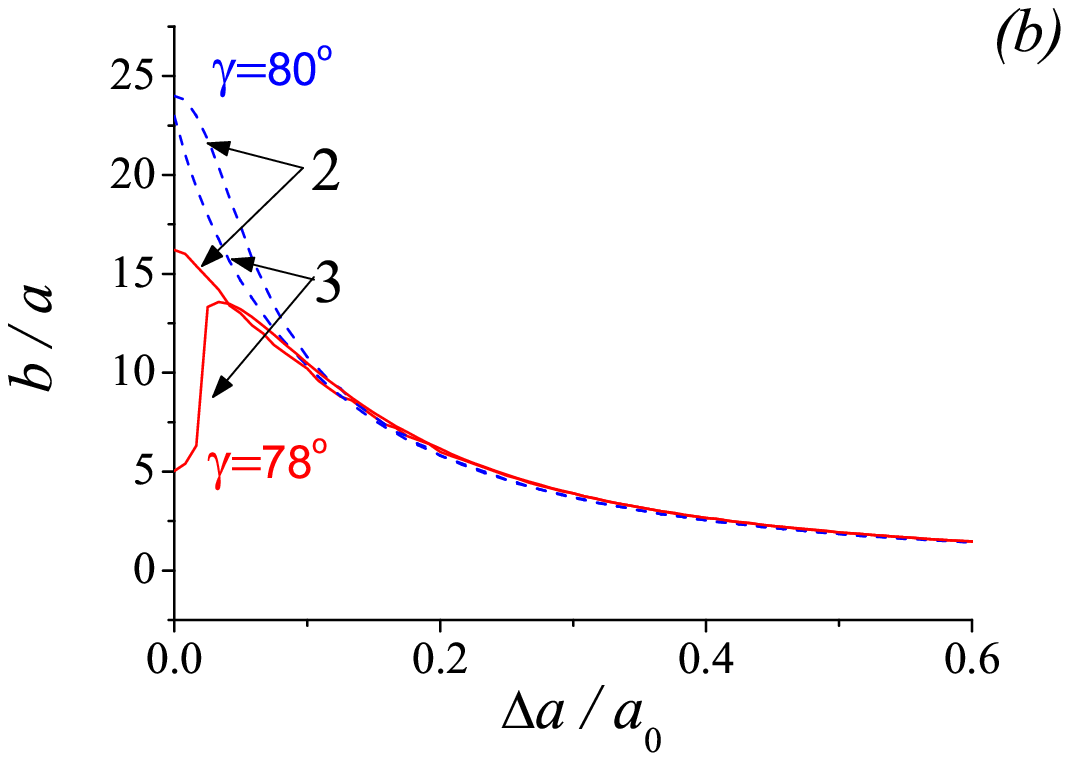}
\caption{ (Color online)(a) The energy difference $\Delta\varepsilon_c$ vs the
relative displacement $\Delta a$ of vortex sublattices for
different tilting angles $\gamma = 78^o$ (solid line) and
$\gamma = 80^o$ (dashed line) and different number of flux
quanta per unit cell $M=2,\, 3$. (b) Lattice deformation ratio
$\sigma=b/a$ vs the relative displacement $\Delta a$ of vortex
sublattices for different tilting angles $\gamma = 78^o$ (solid line) and
$\gamma = 80^o$ (dashed line) and different number
of flux quanta per unit cell $M=2,\, 3$.
Here we put $a_0=60\lambda_{ab}$.
The numbers near the curves denote the number $M$ of vortices per
unit cell.} \label{Fig:11}
\end{figure}


\end{document}